\documentclass[reprint, amsmath,amssymb, aps]{revtex4-2}
\usepackage{graphicx}% Include figure files
\usepackage{dcolumn}% Align table columns on decimal point
\usepackage{bm, hyperref, color, xcolor}
\usepackage{subfigure, comment, float}
\usepackage[normalem]{ulem}

\begin{document}

\preprint{APS/123-QED}

\title{Active sorting to boundaries in active nematic -- passive isotropic fluid mixtures}

\author{Saraswat Bhattacharyya}
 
\author{Julia M. Yeomans}%
\email{julia.yeomans@physics.ox.ac.uk}
 \
\affiliation{%
Rudolf Peierls Centre for Theoretical Physics, Parks Road, University of Oxford, OX1 3PU. United Kingdom.
}%

\date{\today}

\begin{abstract}
We use a two-fluid model to study a confined mixture of an active nematic fluid and a passive isotropic fluid. We find that an extensile active fluid preferentially accumulates at a boundary if the anchoring is planar, whereas its boundary concentration decreases for homeotropic anchoring. These tendencies are reversed if the active fluid is contractile. We argue that the sorting results from gradients in the nematic order, and show that the behaviour can be driven by either imposed boundary anchoring or spontaneous anchoring induced by active flows. Our results can be tested by experiments on microtubule-kinesin motor networks, and may be relevant to sorting to the boundary in cell colonies or cancer spheroids.

\end{abstract}

\maketitle

\section{Introduction}
\label{Section: Intro}

% Phase separation in biology 
A central challenge in biophysics is to understand how numerous individual components self-organize to form more ordered structures. For example,  experiments with starfish embryos in which endodermal and ectodermal cells were mixed showed that the cells spontaneously sorted, with endodermal cells migrating towards the interior and ectodermal cells moving to the outer boundary of the embryo \cite{Suzuki2021}. Similarly, in Huang et al.\cite{huang2020architecture}, a co-culture of homogeneously mixed metastatic breast cancer cells and non-tumorigenic epithelial cells initially sorted and later exhibited an inversion of tissue architecture. In confluent cell layers, different cell types can self-arrange into distinct regions \cite{KRENS2011,Balasubramaniam2021}, and bacterial species within a biofilm have been observed to spontaneously sort based on differences in their mechanical properties \cite{Oldewurtel2015}. 

%DAH and free energy formalism
Biological phase separation has often been examined using the framework of equilibrium thermodynamics, where sorting is assumed to stem from the minimization of a free energy. Among the most prominent of the equilibrium theories is the differential adhesion hypothesis (DAH) \cite{Steinberg1975, Foty2005}, which attributes the ordering to variations in the binding strength between like and unlike cells. However, since living matter operates out of equilibrium, the DAH may not adequately account for factors such as chemical signalling, cell division, and mechanical forces. Indeed, recent experiments have demonstrated that the DAH alone is insufficient to fully explain cell sorting in living systems \cite{Tse2021,Heine2021,Pawlizak2015}.

% Active Matter and Active PS 
A class of systems exhibiting out-of-equilibrium phase separation is active matter \cite{marchetti2013hydrodynamics, Cates2025}, which describes self-motile particles. A paradigmatic example of active phase separation is motility-induced phase separation (MIPS), where a system of self-propelled particles separates into dense and dilute phases \cite{cates2015motility}. Other routes to active phase separation include aligning torques, chemical bonds, motility differences, and hydrodynamic interactions \cite{Zhang2021, Gokhale2022, Thutupalli2018, Furukawa2014,Lauersdorf2024,Beatrici2011,Belmonte2008}. Active phase separation has been documented in various continuum models \cite{Stenhammer2013, Wittkowski2014, Weber2018, Tiribocchi2015}, as well as in phase field and vertex model simulations of heterogeneous cell layers \cite{Balasubramaniam2021, Graham2024, Rozman2024}. The sorting can also be influenced by confining the active material; in many active mixtures, the different species preferentially aggregate at or move away from a boundary \cite{Yang2014,Neville2024,PerezGonzalez2019}. Swimming microorganisms, in particular, congregate at walls by hydrodynamic interactions \cite{Berke2008, Spagnolie2012}.

% Active nematics - generic
A widely used formalism for studying dense cellular aggregates is that of active nematics \cite{doostmohammadi2018active,Saw2017,duclos2018spontaneous}. These comprise elongated particles that align in a common direction but lack positional order. The particles exert local dipolar forces on their surroundings by pushing or pulling along their long axis. 
%Active nematics - confinement and interfaces
A consequence of this is that active nematics in bulk are unstable to a chaotic flow state called active turbulence \cite{SimhaRamaswamy2002}. By contrast, in confinement, they can exhibit a rich variety of steady flow states, including unidirectional and bidirectional laminar flow, spontaneous rotational flows, and a lattice of vortices which can be controlled by adjusting the dimensions  and the boundary conditions of the confining box \cite{voituriez2005spontaneous,shendruk2017dancing,opathalage2019self, Caballero2023}. Moreover, active nematics at a boundary or interface give rise to interesting new physics, notably active anchoring \cite{Blow2014}, active waves \cite{Soni2019, Gulati2024, Adkins2022}, finger formation \cite{Alert2019}, and interfacial self-folding \cite{zhao2024asymmetric}. 

%Boundary effects in other active models % and % Mission statement
We have recently demonstrated that a binary mixture of an active nematic and a passive  fluid  will sort even in the absence of thermodynamic driving forces. However, many of the experiments demonstrating sorting are carried out using confined cell collectives, and the effect of boundaries on the active sorting remains unclear. In this paper we use a two-fluid model to investigate this problem numerically \cite{Bhattacharyya_2023, Weber2018, Joanny2007}. We show that, in the absence of thermodynamically-controlled wetting, whether the active or the passive component preferentially sorts to the boundary depends on the director anchoring, and we give analytical arguments to explain this result.

% Structure overview : Uncomment the paragraph below 
The paper is structured as follows: In Sec.~\ref{sec:Model}, we review the model introduced in Bhattacharyya et al.\cite{Bhattacharyya_2023, bhattacharyya2024phase} to study a mixture of an active nematic and a passive isotropic fluid and give details of the numerical approach. In Sec.~\ref{sec:Box}, we confine the mixture to a square box, imposing either homeotropic or parallel anchoring on the nematic director. We find that an extensile (contractile) active nematic component preferentially sorts to the wall if the anchoring is planar (homeotropic). We then show that the behaviour is similar if the wall anchoring itself results from activity (Sec.~\ref{sec:Active}). 
In Sec.~\ref{sec:Drop} we turn to confinement in a circle. This allows us to stabilise a circulating flow pattern and hence to compare the effects of curvature forces to those resulting from gradients in the magnitude of the nematic ordering. The final section,~\ref{sec:End}, summarises the results and suggests a possible mechanism for the inversion of tissue architecture observed in Huang et al.\cite{huang2020architecture}.

\section{Model}
 \label{sec:Model}

We consider a mixture of two fluid components $i=1,2$ where $i=1$ is an active nematic and $i=2$ is a passive, isotropic fluid. Each component satisfies mass and momentum conservation:
\begin{align}
       \partial_t \rho^{i} + \nabla \cdot( \rho^{i} \mathbf{u}^{i} ) &= 0 \:, \label{eqn:CompMassContinuity} \\
    \partial_t ( \rho^{i} \mathbf{u}^{i}) + \nabla \cdot ( \rho^{i} \mathbf{u}^{i} \mathbf{u}^{i} ) &= \mathbf{f}^{visc, i} + \mathbf{f}^{thermo, i}  + \mathbf{f}^{body, i} \nonumber \\ &\qquad\quad + \gamma \phi (1-\phi) (\mathbf{u}^{3-i} - \mathbf{u}^{i}) \:
    \label{eqn:CompMomentumBalance}
\end{align}
where $\rho^{i}$ and $\mathbf{u}^{i}$ are the density and velocity of component $i$ and $\phi$ is the concentration of component 1. The right hand side of Eq.\eqref{eqn:CompMomentumBalance} describes the forces acting per unit volume of the fluid, and accounts for viscous ($\mathbf{f}^{visc, i}$), thermodynamic ($\mathbf{f}^{thermo, i}$), and local body forces ($\mathbf{f}^{body, i}$). In addition, the term  $\gamma \phi (1-\phi) (\mathbf{u}^{3-i} - \mathbf{u}^{i})$ models relative drag between the two fluid components, the strength of which is controlled by the coefficient $\gamma$.

It is convenient to reformulate the equations of motion in terms of a centre-of-mass (COM) fluid, and a relative fluid. The density field of the combined COM fluid is $\rho^c = \rho^1 + \rho^2$, and the COM flow field is $\mathbf{u}^c = \phi\mathbf{u}^1+(1-\phi)\mathbf{u}^2$. We also define a relative flow field $\delta\mathbf{u} = \mathbf{u}^1 - \mathbf{u}^2$. We will work in the limit where the relative drag is the fastest mode of relaxation in the system i.e.~$\gamma$ is large. Thus, the relative flow is much smaller than the combined velocity of the fluid i.e. $|\delta \mathbf u| \ll |\mathbf u^c|$.
Adding Eqs.\eqref{eqn:CompMassContinuity}-\eqref{eqn:CompMomentumBalance} for each fluid, and dropping terms of order $(\delta\mathbf{u})^2$, gives the equations of motion for the COM fluid \cite{Malevanets1999, Malevanets2000, Bhattacharyya_2023, bhattacharyya2024phase}:
\begin{align}
       \partial_t \rho^{c} + \nabla \cdot( \rho^{c} \mathbf{u}^{c} ) &= 0 \:, \label{eqn:CombiMassContinuity} \\
    \partial_t ( \rho^{c} \mathbf{u}^{c}) + \nabla \cdot ( \rho^{c} \mathbf{u}^{c} \mathbf{u}^{c} ) &= \sum_{i=1}^2 \big[  \mathbf{f}^{visc, i} + \mathbf{f}^{thermo, i}  + \mathbf{f}^{body, i} \big]\:.
    \label{eqn:CombiMomentumBalance}
\end{align}

We now describe the terms on the right-hand side of Eq.\eqref{eqn:CombiMomentumBalance}. The viscous term can be written as the divergence of a viscous stress tensor,
\begin{equation}
     \mathbf{f}^{visc, i} = \mathbf{\nabla}\cdot \, \rho^i \nu^i \big(\,(\nabla\,\mathbf{u}^i) + (\nabla\,\mathbf{u}^i)^T -  (\nabla\cdot\mathbf{u}^i)\,\mathbb{I}\,\big),
    \label{eqn:ViscousStress_comp}
\end{equation}
where $\nu^i$ is the kinematic viscosity of fluid $i$. To simplify things further, we assume that $\nu^1 = \nu^2 = \nu$. Since $|\delta\mathbf{u}|\ll|\mathbf{u}^c|$, the viscous term for the combined fluid is then \cite{Malevanets1999, Malevanets2000}
\begin{equation}
    \mathbf{f}^{visc, c} = \mathbf{\nabla}\cdot\rho^c \nu \,\big(\,(\nabla\,\mathbf{u}^c) + (\nabla\,\mathbf{u}^c)^T -  (\nabla\cdot\mathbf{u}^c)\,\mathbb{I}\,\big).
    \label{eqn:ViscousStress_combi}
\end{equation}

The second term on the right-hand side of Eq.~\eqref{eqn:CombiMomentumBalance} is the thermodynamic force derived from a free energy density $\mathcal{F}$, which we will describe below. The thermodynamic force acting on each fluid component can be written as
\begin{equation}
    \mathbf{f}^{thermo, i} = -\rho^i \mathbf{\nabla}\mu^i\:,
\end{equation}
where $\mu^i = \delta\mathcal{F}/\delta\rho^i$ is the chemical potential of species $i$. The total thermodynamic force acting on the COM fluid can  be written as the divergence of
a stress tensor, $\mathbf{\sigma}^{thermo}$, \cite{Malevanets1999, Bhattacharyya_2023}:
\begin{align}
    \mathbf{f}^{thermo, c} &= \nabla\cdot\mathbf{\sigma}^{thermo} = \nabla\cdot\bigg[\big(  \mathcal F - \frac{\delta \mathcal F}{\delta \rho^1} \rho^1 - \frac{\delta \mathcal F}{\delta \rho^2} \rho^2 \big) \,\mathbb{I \bigg]}.
\end{align}

The final term on the right-hand side of Eq.\eqref{eqn:CombiMomentumBalance} is a body force that models the active and passive forces arising from the nematic nature of the first fluid component. The nematic order is described by a rank-2 nematic tensor field \cite{degennes_book}
\begin{equation}
    \mathbf{Q} = 2S\Big(\mathbf{n}\,\mathbf{n} - \frac{\mathbb{I}}{2}\Big) \:,
\end{equation}
where $S$ is the magnitude and $\mathbf{n}$ is the alignment axis of the nematic ordering. The dynamics of $\mathbf{Q}$ is governed by the Beris-Edwards equation \cite{beris1994thermodynamics, doostmohammadi2018active}
\begin{equation}
    (\partial_t + \mathbf{u}^1\cdot\nabla)\, \mathbf{Q} + \mathbf{W} = \Gamma\mathbf{H} \:,
    \label{eq:BE}
\end{equation}
where 
\begin{align}
	\mathbf{W} &= \left( \lambda \mathbf{\tilde E}^1 + {\Omega}^1 \right) \cdot\Big( \mathbf{Q} + \frac{\mathbb I}{2} \Big) + \Big( \mathbf{Q} + \frac{\mathbb I}{2} \Big) \cdot\left( \lambda \mathbf{\tilde E}^1-\Omega^1 \right) \nonumber \\ & \qquad\qquad\qquad\qquad- 2\lambda \Big( \mathbf{Q} + \frac{\mathbb I}{2} \Big) (\mathbf{Q}:\mathbf{\tilde E}^1)
    \label{eq:2meth:Q_rot}
    \end{align}
is the co-rotation term describing the response of the orientation field to the traceless \cite{Supavit2022} strain rate tensor $\mathbf{\tilde E}^1$ and vorticity tensor $\Omega^1$ in the fluid. The flow-tumbling parameter $\lambda$ describes the relative importance of strain and vorticity. Finally,  
\begin{equation}
    \mathbf{H} = -\left( \frac{\delta \mathcal{F}}{\delta \mathbf{Q}} - \frac{\mathbb{I}}{2} \text{Tr}\left( \frac{\delta \mathcal{F}}{\delta \mathbf{Q}} \right) \right)
    \label{eq:2meth:Mol_field}
\end{equation}
is a molecular field describing relaxation to the minimum of the free energy at a rate depending on $\Gamma$, the rotational diffusivity.

Distortions in the nematic field exert a restoring stress on the underlying fluid, known as the elastic backflow term. This is given by \cite{doostmohammadi2018active} 
\begin{align}
     \Pi^{backflow} &= \phi\Bigg( 2\lambda \Big(\mathbf{Q} + \frac{\mathbb{I}}{2} \Big) (\mathbf{Q : H})-  \lambda \mathbf{H}\cdot \Big( \mathbf{Q} + \frac{\mathbb{I}}{2} \Big)    \nonumber\\
& \kern-1em \kern-1em- \lambda \Big( \mathbf{Q}  + \frac{\mathbb{I}}{2} \Big)\cdot\mathbf{H} -\big( \nabla \mathbf{Q} \big)\, . \frac{\partial \mathcal F}{\partial \nabla \mathbf{Q}} + \mathbf{Q \cdot H} - \mathbf{H \cdot Q} \: \Bigg) .  
     \label{eq:2meth:Pi_el}
\end{align}
Importantly, the active nematic species exerts dipolar forces along the local elongation axis. This can be modelled by an active stress \cite{SimhaRamaswamy2002, doostmohammadi2018active}
\begin{equation}
    \Pi^{act} = -\zeta\phi\mathbf{Q}\:,
    \label{eq:active}
\end{equation}
where the activity parameter $\zeta$ describes the strength of the dipolar forces and the sign of $\zeta$ determines their direction. Positive (negative) $\zeta$ corresponds to extensile (contractile) activity, where the elementary dipoles exert a net force outwards (inwards) along their long axis.
Combining Eqs.(\ref{eq:2meth:Pi_el}) and (\ref{eq:active}), the total body force appearing in Eq.\eqref{eqn:CombiMomentumBalance} is 
\begin{equation}
    \mathbf{f}^{body, i} = \delta_{i,1} \mathbf{\nabla}\cdot(\Pi^{backflow} + \Pi^{act})\:.
\end{equation}

It now remains to define the free energy $\mathcal{F}$ which is used to calculate the chemical potential and the molecular field. We choose \cite{bhattacharyya2024phase}
\begin{align}
    \mathcal{F} &= \frac{1}{3} \rho^c\ln\rho^c + \Big\{ a(\phi-\frac{1}{2})^2 + b(\phi-\frac{1}{2})^4 \Big\} \nonumber \\ & \qquad+\Big[  \frac{1}{2}C \big(S_0^2-\frac{1}{2}\text{Tr}(\mathbf{Q}^2)\big)^2 + \frac{1}{2} K(\mathbf{\nabla}\mathbf{Q})^2 \Big]\:
    \label{eq:FullFreeEnergy}
\end{align}
where the first term defines an isothermal equation of state for the combined fluid. Thus, the combined fluid is essentially incompressible, and density variations are resolved much faster than the timescale of system dynamics. However, the individual component fluids are compressible, allowing sorting.

The second term in Eq.\eqref{eq:FullFreeEnergy}, in curly brackets, is the Landau free energy. The material parameters $a$ and $b$ describe the equilibrium state of the system in the absence of activity. If $a>0$ and $b>0$, equilibrium corresponds to the active nematic and passive isotropic fluids being homogeneously mixed. 

The final term in Eq.\eqref{eq:FullFreeEnergy}, in square brackets, is the Landau-de Gennes free energy describing the thermodynamics of a passive nematic.  $C$ and $K$ are bulk material properties of the nematic, and $S_{0}$ is the magnitude of the nematic order in equilibrium. We take $S_{0}=0$ corresponding to a paranematic, i.e.~no nematic order in the absence of activity or boundaries.\\

We solve Eqs.\eqref{eqn:CombiMassContinuity}--\eqref{eqn:CombiMomentumBalance} for the combined fluid using an incompressible Lattice-Boltzmann method \cite{LB_book}. Eqs.\eqref{eqn:CompMassContinuity}--\eqref{eqn:CompMomentumBalance} for the compressible active component are solved using a compressible Lattice Boltzmann method \cite{Bhattacharyya_2023, Malevanets1999}. Eq.\eqref{eq:BE} for the $\mathbf{Q}$-field is integrated using a finite difference approach. More details of the simulation techniques can be found in Bhattacharyya et al.\cite{Bhattacharyya_2023, bhattacharyya2024phase} where we used the algorithm to study the ordering of bulk mixtures of an active and a passive fluid, and mixtures of active fluids.

% Adding confinement
In this paper, we extend our previous work by confining an active-passive mixture. We consider two different geometries. In the first of these, the mixture is confined inside a square box of side $L$ with solid walls. We impose free-slip boundary conditions for each fluid component  at the box walls. This is simulated numerically by using a Lattice Boltzmann bounce-back scheme. The orientation field is constrained to be perfectly anchored at the walls in either a planar or homeotropic configuration. 

In the second case, the mixture is confined inside a fixed region by a third highly viscous fluid, with no-slip and no-penetration conditions at the boundary. The nematic field at the boundary is free, and its orientation is controlled by active flows.
Numerically, we use a fixed phase field $\Psi$ to distinguish between the inside ($\Psi=1$) and the outside ($\Psi=0$) of the confinement. In order to have a smooth boundary, $\Psi$ is chosen to minimize the free energy $F_\Psi = A_\Psi\Psi^2(1-\Psi)^2 + K_\Psi(\nabla\Psi)^2$ with $A_\Psi=0.4$ and $K_\Psi=0.1$. The external confining fluid is chosen to have the same density as the fluid mixture inside, but it is highly viscous ($\nu^{out} = 10/6$) and has strong frictional damping $\mathbf{f}^{fric, i}_{\Psi=0} = -10\mathbf{u}^i$,  ensuring no flow outside the confinement. 
At the boundary, $\Psi$ varies smoothly from $1$ to $0$ over 4 lattice sites. The activity is localized inside the confined region by setting it to 0 if $\Psi<0.90$.   To deal with the finite width of the interface, we linearly interpolate the kinematic viscosity and friction coefficient for $\Psi \in (0, 1)$. The velocity field is continuous at the interface, and zero outside the boundary. %\sout{, leading to no-slip and no-penetration conditions across the boundary.} %Moreover, the flow field outside the confined region is damped by a strong frictional term ($-b\mathbf{u}$, with $b=0.1$). The inner and outer regions are coupled across the boundary by continuity of stress, and a no-penetration condition.  

%Choice of parameters
We use the parameters $\rho^C = 40$, $\nu_1 = \nu_2 = 1/6$, $\gamma =0.1$, $a=0.0001$, $b=0$, $\Gamma=0.33$, $C=0.001$, $K = 0.02$, $S_0=0$, $\lambda=0.7$, and $\zeta \in [0.001, 0.012]$. The system is initialized in a uniformly mixed configuration, with equal concentrations of active and passive material ($\phi=1/2$) everywhere. We initialize the nematic tensor in a random configuration with noise, and then simulate for $500$ burn-in steps with no activity. Simulations are then run for at least $6\times10^5$ timesteps, with snapshots taken every $1000$ timesteps, until a steady state is reached. 
We quantify the typical concentration and fluctuations of the active fluid near the boundary, $\phi_{edge}$, by measuring the mean and standard deviation of $\phi$ in a small region near the boundary, across different time snapshots. For a circular (square) confinement, these regions are given by annuli (square annuli) of width 2 units. 

\section{Active sorting to boundaries driven by imposed anchoring \label{sec:Box}}

\begin{figure*}[htp]
    \centering
    \includegraphics[width = \textwidth]{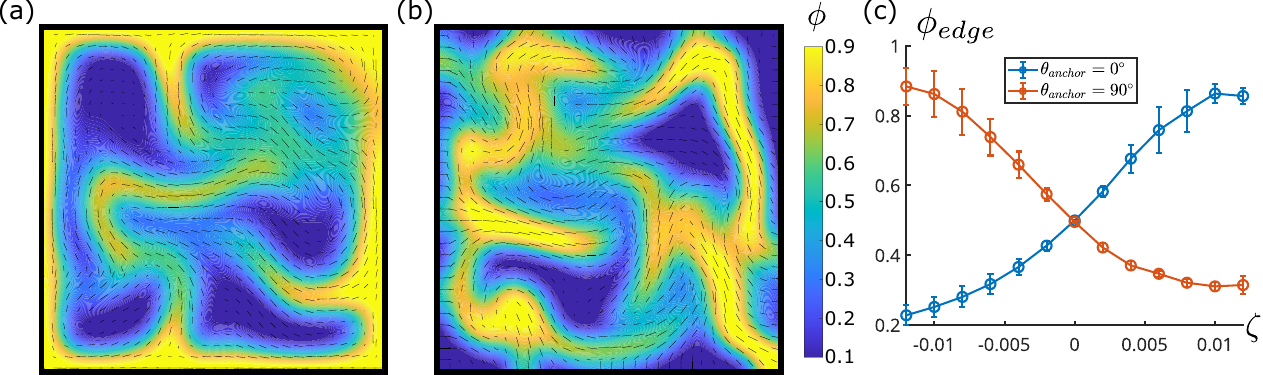}
    \caption{\textbf{Active nematic-passive mixture with imposed strong anchoring:} \textit{(a)} \textbf{Extensile activity, planar anchoring:} For extensile activity ($\zeta=0.01$), with planar anchoring, the concentration of active nematic at the boundary wall is enhanced. The colourbar shows the concentration of active nematic fluid $\phi$.\textit{(b)} \textbf{Extensile activity, homeotropic anchoring:} If the anchoring is changed to homeotropic, the concentration of active nematic at the boundary is depleted. \textit{(c)} Concentration of active nematic near the wall, averaged over time, for varying activity. Error bars show
the standard deviation.} %The active material accumulates at the wall for extensile (contractile) nematics with planar (homeotropic) anchoring, and depletes for the configuration with opposite anchoring. 
    \label{fig:BasicActiveWetting_Box}
\end{figure*}

We first study an equal mixture of an active nematic and a passive isotropic fluid confined inside a square box ($L=120$). The velocity field satisfies a free-slip boundary condition, while the nematic director field at the walls is anchored in a planar (parallel to wall) or homeotropic (perpendicular to wall) configuration. 

In the bulk of the box we observe active turbulence and microphase separation as detailed in Bhattacharyya et al.~\cite{Bhattacharyya_2023}. Note, however that the snapshots in Fig.~\ref{fig:BasicActiveWetting_Box} and Movies 1--2, which are for a mixture with extensile activity $\zeta = 0.01$, suggest that the active nematic component preferentially accumulates at the boundary if the anchoring is planar (panel a) but has a lower concentration at the boundary for homeotropic anchoring (panel b). 

These observations are quantified in Fig.~\ref{fig:BasicActiveWetting_Box}(c) where the concentration of the active fluid component near the wall is plotted as a function of activity for the two anchoring configurations. This figure also allows a comparison of sorting for a contractile active component, where the opposite behaviour is seen, with the wall concentration being suppressed by planar, and enhanced by homeotropic, anchoring. This comparison indicates that the concentration ordering at the boundary is driven by active forces.

\begin{figure}[H]
    \centering
    \includegraphics[width = 0.35\textwidth]{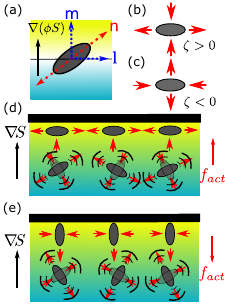}%was 0.47
    \caption{\textbf{Forces at a boundary:}  Schematic representation of \textit{(a)} a nematic at a boundary showing the vectors defined in the text. Background colours show the magnitude of nematic order, with yellow denoting high nematic order and blue denoting low nematic order. The gradient of nematic order points from blue to yellow. \textit{(b)} flows due to an extensile ($\zeta>0$) active nematic. \textit{(c)} flows due to a contractile ($\zeta<0$) active nematic. \textit{(d)-(e)}  nematic order gradient and active forces at an anchored boundary. Extensile nematics are ordered at the boundary, but disordered in the bulk (shown by curved black lines), resulting in a gradient in nematic order, $S$. The resultant active force $f_{act}$ points \textit{(d)} towards the boundary for planar anchoring, and \textit{(e)} away from the boundary for homeotropic anchoring. }
    \label{fig:Mech}
\end{figure}

%Forces at an interface
 Indeed, the boundary sorting can be explained by noting the form of the active stress, Eq.~\eqref{eq:active}, which implies that gradients of nematic order or material concentration generate active forces. Consider a nematic field, with  orientation  $\mathbf{n}$, at a boundary, with unit vectors $\mathbf{m}$ and $\mathbf{l}$ denoting the normal and tangential directions to the boundary, as shown in in Fig.~\ref{fig:Mech}(a). For a gradient of nematic order or concentration $\nabla(\phi S)$ that points along $\mathbf{m}$, the component of the active force normal to the boundary is $F^{norm} = \zeta \,| \nabla (\phi S) |\, (2 (\mathbf{m \cdot n})^2 -1 ) \,\mathbf{m} $ \cite{Blow2014}. Its direction depends on the anchoring angle of the nematogens at the boundary, for example for an extensile nematic with planar anchoring $F^{norm}$ points in the same direction as the gradient,  $\nabla (\phi S)$. 
%\newpage

\begin{figure*}[htp]
    \centering
    \includegraphics[width = \textwidth]{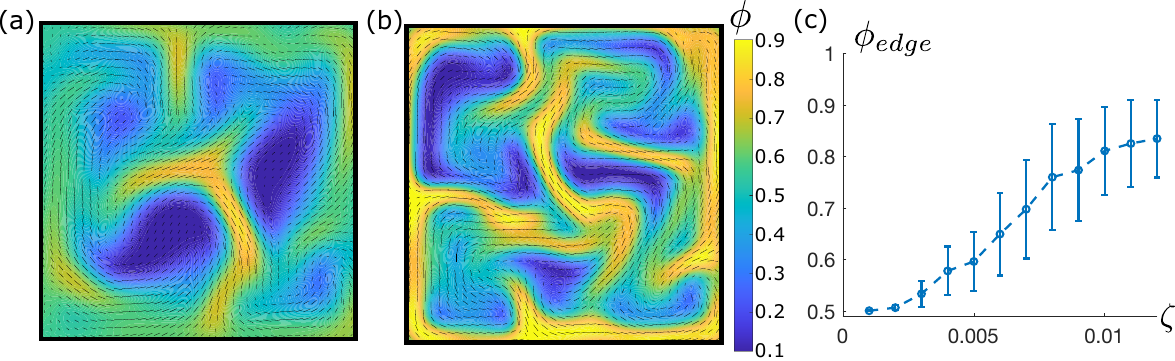}
    \caption{\textbf{Active nematic-passive mixture with active anchoring:} \textit{(a)} \textbf{Extensile activity, active anchoring:} At a low activity, $\zeta=0.004$, the active material has a slightly enhanced concentration at the boundary. \textit{(b)} \textbf{Extensile activity, active anchoring:} At higher activity, $\zeta=0.01$, bulk flows are chaotic, but the active anchoring, and hence the sorting is stronger. The colourbar shows the concentration of active nematic fluid $\phi$. 
    \textit{(c)} Average concentration of active nematic near the wall for varying activity. Error bars show the standard deviation.}
    \label{fig:BasicActiveWetting_Circle}
\end{figure*}

The gradients of $S$ and $\phi$ play a similar role as an interface between two immiscible fluids, and the resultant forces look similar to those described by Coelho et al.~\cite{Coelho2023}.

Thus the accumulation of the active fluid at the anchored boundary can be attributed to the nematic order gradient between the nematogens influenced by the boundary anchoring and the more disordered bulk. For an extensile nematic with planar anchoring, the order gradient sets up active flows towards the boundary (Fig.~\ref{fig:Mech}(b,d)). This increases the concentration of extensile nematic at the boundary, setting up a concentration gradient, $\nabla\phi$, which reinforces the order gradient. The balance between the active force and these restoring forces controls the magnitude of the sorting. The sorting is opposed by the chaotic flows of active turbulence and by the Landau free energy, which favours a mixed state. 

Fig.~\ref{fig:Mech}(e) contrasts the case of homeotropic anchoring, where the gradient-induced flows are away from the wall. The change in behaviour can easily be explained  by noting the change in the sign of the activity, $\zeta$. The flow direction would be reversed for a contractile nematic (Fig.~\ref{fig:Mech}(c)). \\

For extensile systems, the bulk is turbulent, and the velocity field in the bulk is higher than that at the edges (for both planar and homeotropic anchoring). However, the velocity field alignment near the edges is uniform, while it is chaotic in the bulk. (Fig.~A1 in the ESI). Similar results are observed for a box with imposed anchoring and no-slip boundary conditions (see Fig.~A2 in the ESI).

\section{Active sorting to boundaries driven by spontaneous active anchoring \label{sec:Active}} 

We next investigate whether sorting to a boundary can be observed even in the absence of imposed anchoring. We consider a square box containing an active nematic and a passive isotropic fluid component in equal amounts, and delineated by a third highly viscous confining fluid. Boundary conditions on the flow are no-slip, and the nematic field at the box boundary is free, so that its orientation is controlled by active flows. 

The results are summarised in Fig.~\ref{fig:BasicActiveWetting_Circle} for extensile activity and a box length $L$=160 which is sufficiently large that the bulk shows active turbulence. We quantify the degree of sorting by measuring the peak of the average concentration profile of the active fluid near the edge of the box. At low activities, $\zeta=0.004$, the snapshot in Fig.~\ref{fig:BasicActiveWetting_Circle}a (see also Movie 3) shows a slight accumulation of the active component at the boundary. For higher activities, $\zeta=0.01$ (Fig.~\ref{fig:BasicActiveWetting_Circle}b, Movie 4), there is a clear sorting of active fluid to the boundary. Fig.~\ref{fig:BasicActiveWetting_Circle}(c) demonstrates the increase of the boundary accumulation with the activity. While this is significant, it is weaker than that caused by strong thermodynamic anchoring. 

 \begin{figure*}[htp]
    \includegraphics[width = \textwidth]{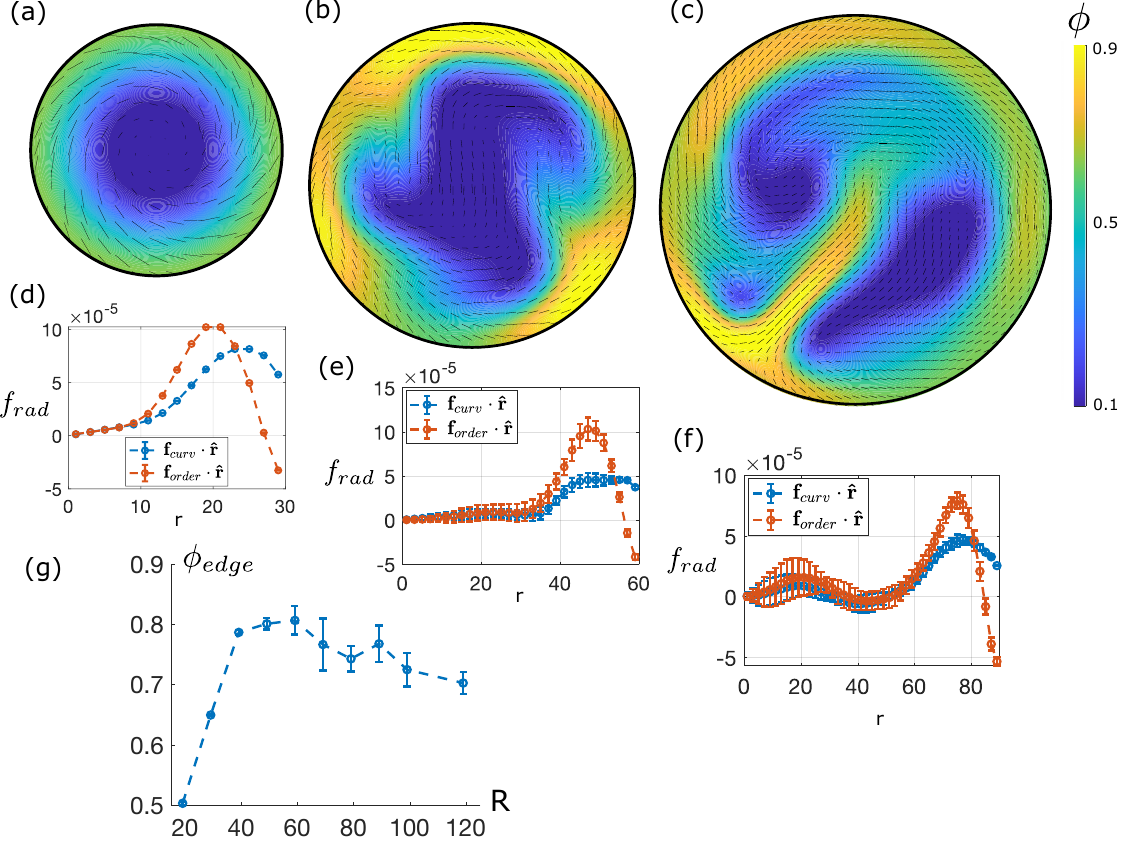}
     \caption{\textbf{Active nematic -- passive mixture in circular confinement:} \textbf{Extensile activity, active anchoring:} Concentration profile of the active nematic fluid, $\phi$ (colourbar) and director (black lines) for \textit{(a)} $R=30$,  \textit{(b)} $R=60$, \textit{(c)} $R=90$ for $\zeta=0.005$.  Comparison of the radial forces due to gradients in the nematic order parameter and gradients in the director field as a function of the radial co-ordinate, $r$, for \textit{(d)} $R=30$, \textit{(e)} $R=60$, \textit{(f)} $R=90$.  \textit{(g)} Average concentration of active nematic near the wall for varying radii of confinement. }
     \label{fig:RotatingActiveDroplets}
\end{figure*}

The boundary sorting that occurs, even in the absence of imposed boundary conditions, is a result of active anchoring, the alignment of nematogens at a boundary caused by active flows \cite{Blow2014, Coelho2021}. Returning to the argument in Section~\ref{sec:Box} and Fig.~\ref{fig:Mech}, the 
component of the active force tangential to the boundary is  $F^{tang} = 2 \zeta\, |\nabla (S\phi)| \,(\mathbf{m \cdot n}) (\mathbf{l \cdot n}) \,\mathbf{l}$ \cite{Blow2014}.  If the director at the boundary is unconstrained, this force creates flows which tend to rotate the orientation field into the planar configuration for extensile fluids. The active anchoring at the surface then acts in the same way as thermodynamic anchoring to establish a gradient of nematic order normal to the surface which leads to preferential accumulation of the active component at the surface. (If $\lambda>0$, there are no active flows for contractile activity as we are considering a system with no nematic order in the passive state and the shear-induced nematic ordering required for active turbulence in this limit requires $\lambda\zeta>0$ \cite{Santhosh2020}.) 

Similar results are observed for a channel with spontaneous active anchoring and no-slip boundary conditions (see Fig.~A3 in the ESI). Figures showing the spatial variation of the active concentration $\phi$ is presented in Fig.~A4 of the ESI.

\section{Active sorting to boundaries driven by correlated director gradients \label{sec:Drop}}

\begin{figure*}[ht]
     \includegraphics[width = \textwidth]{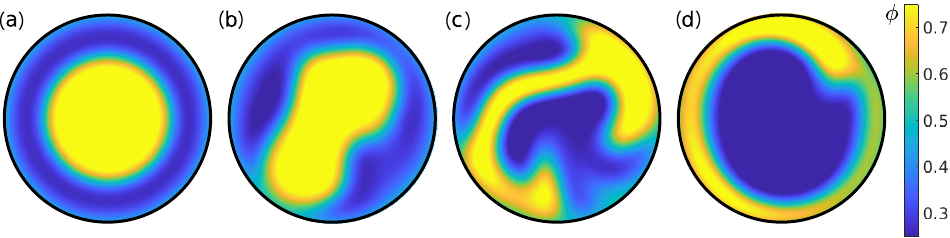}
     \caption{\textbf{Active Sorting:} \textbf{Extensile activity, active anchoring:} \textit{(a)} Stable initial configuration in a thermodynamically phase-separated system ($a<0, b>0$).  \textit{(b)} When the inner species becomes active, it elongates into an elliptical shape. \textit{(c)} On contact with the boundaries, the active material spreads along the confinement. \textit{(d)} The active material accumulates into a ring at the edge of the container, which spontaneously rotates.  The colourbar shows the concentration of the active species, $\phi$. }
     \label{fig:ActiveSorting}
 \end{figure*}
 
Returning to the expression (\ref{eq:active}) for the active stress, we have so far considered the active forces due to gradients in concentration or nematic order, 
\begin{equation}
    \mathbf{F}^{order} = \zeta\mathbf{\nabla}(S\phi)\cdot (2\,\mathbf{n}\,\mathbf{n} - \mathbb{I})\:.
    \label{maggrad}
\end{equation}
However, gradients in the direction of the nematic order parameter may also lead to flows that can push the active fluid component radially,
\begin{equation}
    \mathbf{F}^{curv} = 2\zeta S\phi\,\mathbf{\nabla}\cdot (\mathbf{n}\,\mathbf{n})\:.
    \label{dirgrad}
\end{equation}

To investigate the relative contributions to sorting from gradients in the magnitude or direction of the order parameter, we consider a circular confining region. 
As predicted theoretically in Woodhouse et al.~\cite{Woodhouse2012} and demonstrated experimentally in Opathalage et al.~\cite{opathalage2019self}, 
for a small confining radius active turbulence is replaced by azimuthal flows with the gradients associated with the curvature of the director field driving spontaneous rotation.  This is shown in Fig.~\ref{fig:RotatingActiveDroplets}(a) (see also Movie 5) for $R=30$ and an activity $\zeta =0.005$.
Figs.~\ref{fig:RotatingActiveDroplets}(b,c) (see also Movies 6--7) show similar plots for $R=60$ and $R=90$, where the bulk shows chaotic active flows. 

The variation with $R$ of the concentration enhancement near the edge of the circle is plotted in Fig.~\ref{fig:RotatingActiveDroplets}(g). This increases to $R \sim 60$ and then decreases slightly as the director ordering is disrupted by  the turbulent bulk flows.
The no-slip boundary conditions imply that the maximum shear-induced nematic order is achieved slightly away from the edge of the confinement. Therefore we calculate the concentration enhancement by averaging over an anulus centred on the peak of the radially-averaged active concentration profile.
 
To assess the relative contributions of the nematic ordering and director curvature gradients to the radial force we assume
azimuthal symmetry, $S=S(r)$, $\phi = \phi(r)$, and $\mathbf{n}=(\cos \Theta, \sin \Theta)$ where $\Theta = \theta + \Theta_0$, with $\Theta_0$ a constant tilting angle.  It then follows from Eqns.~(\ref{maggrad}) and~(\ref{dirgrad}) that \cite{Bhattacharyya_2023}
\begin{align}
     \mathbf{f}_{order}\cdot\mathbf{\hat r} &= -\zeta\cos(2\Theta_0) \,\partial_r (S\phi) \:, \\  
     \mathbf{f}_{curv}\cdot\mathbf{\hat r} &= -\zeta\phi\cos(2\Theta_0) \Big(\frac{2S}{r} \Big) \:.
\end{align}
These contributions, which can be calculated numerically, are plotted in Fig.~\ref{fig:RotatingActiveDroplets}(d,e,f) for $R=30,60,90$. We find that  $\mathbf{f}_{order}$ and $\mathbf{f}_{curv}$ contribute approximately equally to setting up the concentration profile. The error bars on these curves indicate that the forces fluctuate much more strongly for larger $R$ because of the active turbulence in the bulk. Note that $\mathbf{f}_{order}\cdot\mathbf{\hat r}$ becomes negative near the boundary - this is a consequence of the active anchoring being suppressed by the no-slip boundary conditions.

\section{Discussion}
\label{sec:End}
 
To summarise, we have shown that a mixture of an active nematic fluid and a passive isotropic fluid spontaneously sorts at a boundary. The sorting mechanism depends on gradients in the nematic order which are set up by boundary anchoring which can be externally imposed or result from active boundary flows. 
 An extensile (contractile) active nematic preferentially accumulates at a boundary where the anchoring is planar (homeotropic) but migrates away from a boundary with homeotropic (planar) anchoring. In the absence of any imposed anchoring, spontaneous anchoring leads to an active species accumulating at the boundaries of its confinement.

Balancing thermodynamic driving with active forces resulting from confinement provides a way to control the positioning of an active fluid component. An example is shown in 
Fig.~\ref{fig:ActiveSorting} (see also Movie 8). A fluid component is initiated at the centre of a confining region, stabilised by thermodynamic phase separation (by choosing $a<0$, $b>0$ %\sar{$a=-2\times10^{-4}$, $b=1.6\times10^{-3}$ , $\zeta=0.004$ if needed }
in Eq.\eqref{eq:FullFreeEnergy}). If the inner fluid becomes active it elongates into an elliptical shape, driven by the active instability, and then spreads into a rotating ring at the boundary. Indeed, cells sorting to the boundary of a confined region or droplet is ubiquitous in biology, and has been observed in  embryonic amphibian cells \cite{Townes1955amphibian}, mouse embryoid cells \cite{Moore2009}, regenerating hydra \cite{cochetEscartin2017hydra}, reconstituted starfish embryo \cite{Suzuki2021}, and cancerous tumours \cite{batlle2012molecular, huang2020architecture}.  Our results highlight a possible role of active forces in driving the cellular organisation, which may be particularly relevant when cells organize opposite to the DAH prediction. For instance, architecture inversion in Huang et al.\cite{huang2020architecture} would be consistent with active nematic stresses associated with cell division, while cell sorting outcomes in Moore et al. \cite{Moore2009} may be driven by active nematic stresses generated by E-cadherin \cite{Balasubramaniam2021}. 

In future work, it would be interesting to study the effects of confinements with locally changing curvatures, and how this affects the boundary distribution of active material.  It might be possible to design confinement shapes such that an active nematic material would spontaneously accumulate in a specific region. Another interesting avenue for exploration would be the evolution of an three-dimensional active droplet with a deformable shape which contains an active-passive mixture. The feedback loop between active flows deforming the shape of the droplet and boundary sorting changing the activity distribution may provide a route to the spontaneous organization of complex patterns.

\section*{Data availability}
The code for the simulations can be found at \url{https://github.com/saraswatb/BoundarySorting}.

\section*{Acknowledgements}

We thank D Cropper,  I Hadjifrangiskou, A Mietke, J Rozman, S Thampi and K Thijssen for useful discussions and feedback.

This research was supported in part by grant no. NSF PHY-2309135 to the Kavli Institute for Theoretical Physics (KITP). SB acknowledges funding from the Rudolf Peierls Centre for Theoretical Physics, and the Crewe Graduate Award. JMY acknowledges support from the ERC Advanced Grant ActBio (funded as UKRI Frontier Research Grant EP/Y033981/1).

\bibliography{rsc}

%apsrev4-2.bst 2019-01-14 (MD) hand-edited version of apsrev4-1.bst
%Control: key (0)
%Control: author (72) initials jnrlst
%Control: editor formatted (1) identically to author
%Control: production of article title (-1) disabled
%Control: page (0) single
%Control: year (1) truncated
%Control: production of eprint (0) enabled
\providecommand{\noopsort}[1]{}\providecommand{\singleletter}[1]{#1}%
\begin{thebibliography}{62}%
\makeatletter
\providecommand \@ifxundefined [1]{%
 \@ifx{#1\undefined}
}%
\providecommand \@ifnum [1]{%
 \ifnum #1\expandafter \@firstoftwo
 \else \expandafter \@secondoftwo
 \fi
}%
\providecommand \@ifx [1]{%
 \ifx #1\expandafter \@firstoftwo
 \else \expandafter \@secondoftwo
 \fi
}%
\providecommand \natexlab [1]{#1}%
\providecommand \enquote  [1]{``#1''}%
\providecommand \bibnamefont  [1]{#1}%
\providecommand \bibfnamefont [1]{#1}%
\providecommand \citenamefont [1]{#1}%
\providecommand \href@noop [0]{\@secondoftwo}%
\providecommand \href [0]{\begingroup \@sanitize@url \@href}%
\providecommand \@href[1]{\@@startlink{#1}\@@href}%
\providecommand \@@href[1]{\endgroup#1\@@endlink}%
\providecommand \@sanitize@url [0]{\catcode `\\12\catcode `\$12\catcode
  `\&12\catcode `\#12\catcode `\^12\catcode `\_12\catcode `\%12\relax}%
\providecommand \@@startlink[1]{}%
\providecommand \@@endlink[0]{}%
\providecommand \url  [0]{\begingroup\@sanitize@url \@url }%
\providecommand \@url [1]{\endgroup\@href {#1}{\urlprefix }}%
\providecommand \urlprefix  [0]{URL }%
\providecommand \Eprint [0]{\href }%
\providecommand \doibase [0]{https://doi.org/}%
\providecommand \selectlanguage [0]{\@gobble}%
\providecommand \bibinfo  [0]{\@secondoftwo}%
\providecommand \bibfield  [0]{\@secondoftwo}%
\providecommand \translation [1]{[#1]}%
\providecommand \BibitemOpen [0]{}%
\providecommand \bibitemStop [0]{}%
\providecommand \bibitemNoStop [0]{.\EOS\space}%
\providecommand \EOS [0]{\spacefactor3000\relax}%
\providecommand \BibitemShut  [1]{\csname bibitem#1\endcsname}%
\let\auto@bib@innerbib\@empty
%</preamble>
\bibitem [{\citenamefont {Suzuki}\ \emph {et~al.}(2021)\citenamefont {Suzuki},
  \citenamefont {Omori}, \citenamefont {Kuraishi},\ and\ \citenamefont
  {Kaneko}}]{Suzuki2021}%
  \BibitemOpen
  \bibfield  {author} {\bibinfo {author} {\bibfnamefont {S.}~\bibnamefont
  {Suzuki}}, \bibinfo {author} {\bibfnamefont {I.}~\bibnamefont {Omori}},
  \bibinfo {author} {\bibfnamefont {R.}~\bibnamefont {Kuraishi}},\ and\
  \bibinfo {author} {\bibfnamefont {H.}~\bibnamefont {Kaneko}},\ }\href
  {https://doi.org/https://doi.org/10.1111/dgd.12749} {\bibfield  {journal}
  {\bibinfo  {journal} {Development, Growth \& Differentiation}\ }\textbf
  {\bibinfo {volume} {63}},\ \bibinfo {pages} {343} (\bibinfo {year}
  {2021})}\BibitemShut {NoStop}%
\bibitem [{\citenamefont {Huang}\ \emph {et~al.}(2020)\citenamefont {Huang},
  \citenamefont {Shiau}, \citenamefont {Wu}, \citenamefont {Segall},\ and\
  \citenamefont {Wu}}]{huang2020architecture}%
  \BibitemOpen
  \bibfield  {author} {\bibinfo {author} {\bibfnamefont {Y.~L.}\ \bibnamefont
  {Huang}}, \bibinfo {author} {\bibfnamefont {C.}~\bibnamefont {Shiau}},
  \bibinfo {author} {\bibfnamefont {C.}~\bibnamefont {Wu}}, \bibinfo {author}
  {\bibfnamefont {J.~E.}\ \bibnamefont {Segall}},\ and\ \bibinfo {author}
  {\bibfnamefont {M.}~\bibnamefont {Wu}},\ }\href@noop {} {\bibfield  {journal}
  {\bibinfo  {journal} {Biophysical Reviews and Letters}\ }\textbf {\bibinfo
  {volume} {15}},\ \bibinfo {pages} {131} (\bibinfo {year} {2020})}\BibitemShut
  {NoStop}%
\bibitem [{\citenamefont {Krens}\ and\ \citenamefont
  {Heisenberg}(2011)}]{KRENS2011}%
  \BibitemOpen
  \bibfield  {author} {\bibinfo {author} {\bibfnamefont {S.~F.~G.}\
  \bibnamefont {Krens}}\ and\ \bibinfo {author} {\bibfnamefont {C.-P.}\
  \bibnamefont {Heisenberg}},\ }\href
  {https://doi.org/https://doi.org/10.1016/B978-0-12-385065-2.00006-2}
  {\bibfield  {journal} {\bibinfo  {journal} {Current Topics in Developmental
  Biology}\ }\textbf {\bibinfo {volume} {95}},\ \bibinfo {pages} {189}
  (\bibinfo {year} {2011})}\BibitemShut {NoStop}%
\bibitem [{\citenamefont {Balasubramaniam}\ \emph {et~al.}(2021)\citenamefont
  {Balasubramaniam}, \citenamefont {Doostmohammadi}, \citenamefont {Saw},
  \citenamefont {Narayana}, \citenamefont {Mueller}, \citenamefont {Dang},
  \citenamefont {Thomas}, \citenamefont {Gupta}, \citenamefont {Sonam},
  \citenamefont {Yap}, \citenamefont {Toyama}, \citenamefont {M{\`e}ge},
  \citenamefont {Yeomans},\ and\ \citenamefont {Ladoux}}]{Balasubramaniam2021}%
  \BibitemOpen
  \bibfield  {author} {\bibinfo {author} {\bibfnamefont {L.}~\bibnamefont
  {Balasubramaniam}}, \bibinfo {author} {\bibfnamefont {A.}~\bibnamefont
  {Doostmohammadi}}, \bibinfo {author} {\bibfnamefont {T.~B.}\ \bibnamefont
  {Saw}}, \bibinfo {author} {\bibfnamefont {G.~H. N.~S.}\ \bibnamefont
  {Narayana}}, \bibinfo {author} {\bibfnamefont {R.}~\bibnamefont {Mueller}},
  \bibinfo {author} {\bibfnamefont {T.}~\bibnamefont {Dang}}, \bibinfo {author}
  {\bibfnamefont {M.}~\bibnamefont {Thomas}}, \bibinfo {author} {\bibfnamefont
  {S.}~\bibnamefont {Gupta}}, \bibinfo {author} {\bibfnamefont
  {S.}~\bibnamefont {Sonam}}, \bibinfo {author} {\bibfnamefont {A.~S.}\
  \bibnamefont {Yap}}, \bibinfo {author} {\bibfnamefont {Y.}~\bibnamefont
  {Toyama}}, \bibinfo {author} {\bibfnamefont {R.-M.}\ \bibnamefont
  {M{\`e}ge}}, \bibinfo {author} {\bibfnamefont {J.~M.}\ \bibnamefont
  {Yeomans}},\ and\ \bibinfo {author} {\bibfnamefont {B.}~\bibnamefont
  {Ladoux}},\ }\href@noop {} {\bibfield  {journal} {\bibinfo  {journal} {Nature
  Materials}\ }\textbf {\bibinfo {volume} {20}},\ \bibinfo {pages} {1156}
  (\bibinfo {year} {2021})}\BibitemShut {NoStop}%
\bibitem [{\citenamefont {Oldewurtel}\ \emph {et~al.}(2015)\citenamefont
  {Oldewurtel}, \citenamefont {Kouzel}, \citenamefont {Dewenter}, \citenamefont
  {Henseler},\ and\ \citenamefont {Maier}}]{Oldewurtel2015}%
  \BibitemOpen
  \bibfield  {author} {\bibinfo {author} {\bibfnamefont {E.~R.}\ \bibnamefont
  {Oldewurtel}}, \bibinfo {author} {\bibfnamefont {N.}~\bibnamefont {Kouzel}},
  \bibinfo {author} {\bibfnamefont {L.}~\bibnamefont {Dewenter}}, \bibinfo
  {author} {\bibfnamefont {K.}~\bibnamefont {Henseler}},\ and\ \bibinfo
  {author} {\bibfnamefont {B.}~\bibnamefont {Maier}},\ }\href
  {https://doi.org/10.7554/eLife.10811} {\bibfield  {journal} {\bibinfo
  {journal} {eLife}\ }\textbf {\bibinfo {volume} {4}},\ \bibinfo {pages}
  {e10811} (\bibinfo {year} {2015})}\BibitemShut {NoStop}%
\bibitem [{\citenamefont {Steinberg}(1975)}]{Steinberg1975}%
  \BibitemOpen
  \bibfield  {author} {\bibinfo {author} {\bibfnamefont {M.~S.}\ \bibnamefont
  {Steinberg}},\ }\href
  {https://doi.org/https://doi.org/10.1016/S0022-5193(75)80091-9} {\bibfield
  {journal} {\bibinfo  {journal} {Journal of Theoretical Biology}\ }\textbf
  {\bibinfo {volume} {55}},\ \bibinfo {pages} {431} (\bibinfo {year}
  {1975})}\BibitemShut {NoStop}%
\bibitem [{\citenamefont {Foty}\ and\ \citenamefont
  {Steinberg}(2005)}]{Foty2005}%
  \BibitemOpen
  \bibfield  {author} {\bibinfo {author} {\bibfnamefont {R.~A.}\ \bibnamefont
  {Foty}}\ and\ \bibinfo {author} {\bibfnamefont {M.~S.}\ \bibnamefont
  {Steinberg}},\ }\href {https://doi.org/10.1016/j.ydbio.2004.11.012}
  {\bibfield  {journal} {\bibinfo  {journal} {Developmental Biology}\ }\textbf
  {\bibinfo {volume} {278}},\ \bibinfo {pages} {255} (\bibinfo {year}
  {2005})}\BibitemShut {NoStop}%
\bibitem [{\citenamefont {Tse}\ \emph {et~al.}(2021)\citenamefont {Tse},
  \citenamefont {Moore}, \citenamefont {Meng}, \citenamefont {Tao},
  \citenamefont {Smith},\ and\ \citenamefont {Xu}}]{Tse2021}%
  \BibitemOpen
  \bibfield  {author} {\bibinfo {author} {\bibfnamefont {J.~D.}\ \bibnamefont
  {Tse}}, \bibinfo {author} {\bibfnamefont {R.}~\bibnamefont {Moore}}, \bibinfo
  {author} {\bibfnamefont {Y.}~\bibnamefont {Meng}}, \bibinfo {author}
  {\bibfnamefont {W.}~\bibnamefont {Tao}}, \bibinfo {author} {\bibfnamefont
  {E.~R.}\ \bibnamefont {Smith}},\ and\ \bibinfo {author} {\bibfnamefont
  {X.-X.}\ \bibnamefont {Xu}},\ }\href
  {https://doi.org/10.1186/s12861-020-00234-0} {\bibfield  {journal} {\bibinfo
  {journal} {BMC Developmental Biology}\ }\textbf {\bibinfo {volume} {21}},\
  \bibinfo {pages} {2} (\bibinfo {year} {2021})}\BibitemShut {NoStop}%
\bibitem [{\citenamefont {Heine}\ \emph {et~al.}(2021)\citenamefont {Heine},
  \citenamefont {Lippoldt}, \citenamefont {Reddy}, \citenamefont {Katira},\
  and\ \citenamefont {Käs}}]{Heine2021}%
  \BibitemOpen
  \bibfield  {author} {\bibinfo {author} {\bibfnamefont {P.}~\bibnamefont
  {Heine}}, \bibinfo {author} {\bibfnamefont {J.}~\bibnamefont {Lippoldt}},
  \bibinfo {author} {\bibfnamefont {G.~A.}\ \bibnamefont {Reddy}}, \bibinfo
  {author} {\bibfnamefont {P.}~\bibnamefont {Katira}},\ and\ \bibinfo {author}
  {\bibfnamefont {J.~A.}\ \bibnamefont {Käs}},\ }\href
  {https://doi.org/10.1088/1367-2630/abf273} {\bibfield  {journal} {\bibinfo
  {journal} {New Journal of Physics}\ }\textbf {\bibinfo {volume} {23}},\
  \bibinfo {pages} {043034} (\bibinfo {year} {2021})}\BibitemShut {NoStop}%
\bibitem [{\citenamefont {Pawlizak}\ \emph {et~al.}(2015)\citenamefont
  {Pawlizak}, \citenamefont {Fritsch}, \citenamefont {Grosser}, \citenamefont
  {Ahrens}, \citenamefont {Thalheim}, \citenamefont {Riedel}, \citenamefont
  {Kießling}, \citenamefont {Oswald}, \citenamefont {Zink}, \citenamefont
  {Manning},\ and\ \citenamefont {Käs}}]{Pawlizak2015}%
  \BibitemOpen
  \bibfield  {author} {\bibinfo {author} {\bibfnamefont {S.}~\bibnamefont
  {Pawlizak}}, \bibinfo {author} {\bibfnamefont {A.~W.}\ \bibnamefont
  {Fritsch}}, \bibinfo {author} {\bibfnamefont {S.}~\bibnamefont {Grosser}},
  \bibinfo {author} {\bibfnamefont {D.}~\bibnamefont {Ahrens}}, \bibinfo
  {author} {\bibfnamefont {T.}~\bibnamefont {Thalheim}}, \bibinfo {author}
  {\bibfnamefont {S.}~\bibnamefont {Riedel}}, \bibinfo {author} {\bibfnamefont
  {T.~R.}\ \bibnamefont {Kießling}}, \bibinfo {author} {\bibfnamefont
  {L.}~\bibnamefont {Oswald}}, \bibinfo {author} {\bibfnamefont
  {M.}~\bibnamefont {Zink}}, \bibinfo {author} {\bibfnamefont {M.~L.}\
  \bibnamefont {Manning}},\ and\ \bibinfo {author} {\bibfnamefont {J.~A.}\
  \bibnamefont {Käs}},\ }\href {https://doi.org/10.1088/1367-2630/17/8/083049}
  {\bibfield  {journal} {\bibinfo  {journal} {New Journal of Physics}\ }\textbf
  {\bibinfo {volume} {17}},\ \bibinfo {pages} {083049} (\bibinfo {year}
  {2015})}\BibitemShut {NoStop}%
\bibitem [{\citenamefont {Marchetti}\ \emph {et~al.}(2013)\citenamefont
  {Marchetti}, \citenamefont {Joanny}, \citenamefont {Ramaswamy}, \citenamefont
  {Liverpool}, \citenamefont {Prost}, \citenamefont {Rao},\ and\ \citenamefont
  {Simha}}]{marchetti2013hydrodynamics}%
  \BibitemOpen
  \bibfield  {author} {\bibinfo {author} {\bibfnamefont {M.~C.}\ \bibnamefont
  {Marchetti}}, \bibinfo {author} {\bibfnamefont {J.-F.}\ \bibnamefont
  {Joanny}}, \bibinfo {author} {\bibfnamefont {S.}~\bibnamefont {Ramaswamy}},
  \bibinfo {author} {\bibfnamefont {T.~B.}\ \bibnamefont {Liverpool}}, \bibinfo
  {author} {\bibfnamefont {J.}~\bibnamefont {Prost}}, \bibinfo {author}
  {\bibfnamefont {M.}~\bibnamefont {Rao}},\ and\ \bibinfo {author}
  {\bibfnamefont {R.~A.}\ \bibnamefont {Simha}},\ }\href@noop {} {\bibfield
  {journal} {\bibinfo  {journal} {Reviews of Modern Physics}\ }\textbf
  {\bibinfo {volume} {85}},\ \bibinfo {pages} {1143} (\bibinfo {year}
  {2013})}\BibitemShut {NoStop}%
\bibitem [{\citenamefont {Cates}\ and\ \citenamefont
  {Nardini}(2025)}]{Cates2025}%
  \BibitemOpen
  \bibfield  {author} {\bibinfo {author} {\bibfnamefont {M.~E.}\ \bibnamefont
  {Cates}}\ and\ \bibinfo {author} {\bibfnamefont {C.}~\bibnamefont
  {Nardini}},\ }\href {https://doi.org/10.1088/1361-6633/add278} {\bibfield
  {journal} {\bibinfo  {journal} {Reports on Progress in Physics}\ }\textbf
  {\bibinfo {volume} {88}},\ \bibinfo {pages} {056601} (\bibinfo {year}
  {2025})}\BibitemShut {NoStop}%
\bibitem [{\citenamefont {Cates}\ and\ \citenamefont
  {Tailleur}(2015)}]{cates2015motility}%
  \BibitemOpen
  \bibfield  {author} {\bibinfo {author} {\bibfnamefont {M.~E.}\ \bibnamefont
  {Cates}}\ and\ \bibinfo {author} {\bibfnamefont {J.}~\bibnamefont
  {Tailleur}},\ }\href@noop {} {\bibfield  {journal} {\bibinfo  {journal}
  {Annual Review of Condensed Matter Physics}\ }\textbf {\bibinfo {volume}
  {6}},\ \bibinfo {pages} {219} (\bibinfo {year} {2015})}\BibitemShut {NoStop}%
\bibitem [{\citenamefont {Zhang}\ \emph {et~al.}(2021)\citenamefont {Zhang},
  \citenamefont {Alert}, \citenamefont {Yan}, \citenamefont {Wingreen},\ and\
  \citenamefont {Granick}}]{Zhang2021}%
  \BibitemOpen
  \bibfield  {author} {\bibinfo {author} {\bibfnamefont {J.}~\bibnamefont
  {Zhang}}, \bibinfo {author} {\bibfnamefont {R.}~\bibnamefont {Alert}},
  \bibinfo {author} {\bibfnamefont {J.}~\bibnamefont {Yan}}, \bibinfo {author}
  {\bibfnamefont {N.~S.}\ \bibnamefont {Wingreen}},\ and\ \bibinfo {author}
  {\bibfnamefont {S.}~\bibnamefont {Granick}},\ }\href
  {https://doi.org/10.1038/s41567-021-01238-8} {\bibfield  {journal} {\bibinfo
  {journal} {Nature Physics}\ }\textbf {\bibinfo {volume} {17}},\ \bibinfo
  {pages} {961} (\bibinfo {year} {2021})}\BibitemShut {NoStop}%
\bibitem [{\citenamefont {Gokhale}\ \emph {et~al.}(2022)\citenamefont
  {Gokhale}, \citenamefont {Li}, \citenamefont {Solon}, \citenamefont {Gore},\
  and\ \citenamefont {Fakhri}}]{Gokhale2022}%
  \BibitemOpen
  \bibfield  {author} {\bibinfo {author} {\bibfnamefont {S.}~\bibnamefont
  {Gokhale}}, \bibinfo {author} {\bibfnamefont {J.~A.}\ \bibnamefont {Li}},
  \bibinfo {author} {\bibfnamefont {A.}~\bibnamefont {Solon}}, \bibinfo
  {author} {\bibfnamefont {J.}~\bibnamefont {Gore}},\ and\ \bibinfo {author}
  {\bibfnamefont {N.}~\bibnamefont {Fakhri}},\ }\href
  {https://doi.org/10.1103/PhysRevE.105.054605} {\bibfield  {journal} {\bibinfo
   {journal} {Physical Review E}\ }\textbf {\bibinfo {volume} {105}},\ \bibinfo
  {pages} {054605} (\bibinfo {year} {2022})}\BibitemShut {NoStop}%
\bibitem [{\citenamefont {Thutupalli}\ \emph {et~al.}(2018)\citenamefont
  {Thutupalli}, \citenamefont {Geyer}, \citenamefont {Singh}, \citenamefont
  {Adhikari},\ and\ \citenamefont {Stone}}]{Thutupalli2018}%
  \BibitemOpen
  \bibfield  {author} {\bibinfo {author} {\bibfnamefont {S.}~\bibnamefont
  {Thutupalli}}, \bibinfo {author} {\bibfnamefont {D.}~\bibnamefont {Geyer}},
  \bibinfo {author} {\bibfnamefont {R.}~\bibnamefont {Singh}}, \bibinfo
  {author} {\bibfnamefont {R.}~\bibnamefont {Adhikari}},\ and\ \bibinfo
  {author} {\bibfnamefont {H.~A.}\ \bibnamefont {Stone}},\ }\href
  {https://doi.org/10.1073/pnas.1718807115} {\bibfield  {journal} {\bibinfo
  {journal} {Proceedings of the National Academy of Sciences of the United
  States of America}\ }\textbf {\bibinfo {volume} {115}},\ \bibinfo {pages}
  {5403} (\bibinfo {year} {2018})}\BibitemShut {NoStop}%
\bibitem [{\citenamefont {Furukawa}\ \emph {et~al.}(2014)\citenamefont
  {Furukawa}, \citenamefont {Marenduzzo},\ and\ \citenamefont
  {Cates}}]{Furukawa2014}%
  \BibitemOpen
  \bibfield  {author} {\bibinfo {author} {\bibfnamefont {A.}~\bibnamefont
  {Furukawa}}, \bibinfo {author} {\bibfnamefont {D.}~\bibnamefont
  {Marenduzzo}},\ and\ \bibinfo {author} {\bibfnamefont {M.~E.}\ \bibnamefont
  {Cates}},\ }\href {https://doi.org/10.1103/PhysRevE.90.022303} {\bibfield
  {journal} {\bibinfo  {journal} {Physical Review E}\ }\textbf {\bibinfo
  {volume} {90}},\ \bibinfo {pages} {022303} (\bibinfo {year}
  {2014})}\BibitemShut {NoStop}%
\bibitem [{\citenamefont {Lauersdorf}\ \emph {et~al.}(2024)\citenamefont
  {Lauersdorf}, \citenamefont {Nazockdast},\ and\ \citenamefont
  {Klotsa}}]{Lauersdorf2024}%
  \BibitemOpen
  \bibfield  {author} {\bibinfo {author} {\bibfnamefont {N.~J.}\ \bibnamefont
  {Lauersdorf}}, \bibinfo {author} {\bibfnamefont {E.}~\bibnamefont
  {Nazockdast}},\ and\ \bibinfo {author} {\bibfnamefont {D.}~\bibnamefont
  {Klotsa}},\ }\href@noop {} {\bibfield  {journal} {\bibinfo  {journal} {arXiv
  preprint arXiv:2407.07826}\ } (\bibinfo {year} {2024})}\BibitemShut {NoStop}%
\bibitem [{\citenamefont {Beatrici}\ and\ \citenamefont
  {Brunnet}(2011)}]{Beatrici2011}%
  \BibitemOpen
  \bibfield  {author} {\bibinfo {author} {\bibfnamefont {C.~P.}\ \bibnamefont
  {Beatrici}}\ and\ \bibinfo {author} {\bibfnamefont {L.~G.}\ \bibnamefont
  {Brunnet}},\ }\href {https://doi.org/10.1103/PhysRevE.84.031927} {\bibfield
  {journal} {\bibinfo  {journal} {Physical Review E}\ }\textbf {\bibinfo
  {volume} {84}},\ \bibinfo {pages} {031927} (\bibinfo {year}
  {2011})}\BibitemShut {NoStop}%
\bibitem [{\citenamefont {Belmonte}\ \emph {et~al.}(2008)\citenamefont
  {Belmonte}, \citenamefont {Thomas}, \citenamefont {Brunnet}, \citenamefont
  {de~Almeida},\ and\ \citenamefont {Chat\'e}}]{Belmonte2008}%
  \BibitemOpen
  \bibfield  {author} {\bibinfo {author} {\bibfnamefont {J.~M.}\ \bibnamefont
  {Belmonte}}, \bibinfo {author} {\bibfnamefont {G.~L.}\ \bibnamefont
  {Thomas}}, \bibinfo {author} {\bibfnamefont {L.~G.}\ \bibnamefont {Brunnet}},
  \bibinfo {author} {\bibfnamefont {R.~M.~C.}\ \bibnamefont {de~Almeida}},\
  and\ \bibinfo {author} {\bibfnamefont {H.}~\bibnamefont {Chat\'e}},\ }\href
  {https://doi.org/10.1103/PhysRevLett.100.248702} {\bibfield  {journal}
  {\bibinfo  {journal} {Physical Review Letters}\ }\textbf {\bibinfo {volume}
  {100}},\ \bibinfo {pages} {248702} (\bibinfo {year} {2008})}\BibitemShut
  {NoStop}%
\bibitem [{\citenamefont {Stenhammar}\ \emph {et~al.}(2013)\citenamefont
  {Stenhammar}, \citenamefont {Tiribocchi}, \citenamefont {Allen},
  \citenamefont {Marenduzzo},\ and\ \citenamefont {Cates}}]{Stenhammer2013}%
  \BibitemOpen
  \bibfield  {author} {\bibinfo {author} {\bibfnamefont {J.}~\bibnamefont
  {Stenhammar}}, \bibinfo {author} {\bibfnamefont {A.}~\bibnamefont
  {Tiribocchi}}, \bibinfo {author} {\bibfnamefont {R.~J.}\ \bibnamefont
  {Allen}}, \bibinfo {author} {\bibfnamefont {D.}~\bibnamefont {Marenduzzo}},\
  and\ \bibinfo {author} {\bibfnamefont {M.~E.}\ \bibnamefont {Cates}},\ }\href
  {https://doi.org/10.1103/PhysRevLett.111.145702} {\bibfield  {journal}
  {\bibinfo  {journal} {Physical Review Letters}\ }\textbf {\bibinfo {volume}
  {111}},\ \bibinfo {pages} {145702} (\bibinfo {year} {2013})}\BibitemShut
  {NoStop}%
\bibitem [{\citenamefont {Wittkowski}\ \emph {et~al.}(2014)\citenamefont
  {Wittkowski}, \citenamefont {Tiribocchi}, \citenamefont {Stenhammar},
  \citenamefont {Allen}, \citenamefont {Marenduzzo},\ and\ \citenamefont
  {Cates}}]{Wittkowski2014}%
  \BibitemOpen
  \bibfield  {author} {\bibinfo {author} {\bibfnamefont {R.}~\bibnamefont
  {Wittkowski}}, \bibinfo {author} {\bibfnamefont {A.}~\bibnamefont
  {Tiribocchi}}, \bibinfo {author} {\bibfnamefont {J.}~\bibnamefont
  {Stenhammar}}, \bibinfo {author} {\bibfnamefont {R.~J.}\ \bibnamefont
  {Allen}}, \bibinfo {author} {\bibfnamefont {D.}~\bibnamefont {Marenduzzo}},\
  and\ \bibinfo {author} {\bibfnamefont {M.~E.}\ \bibnamefont {Cates}},\ }\href
  {https://doi.org/10.1038/ncomms5351} {\bibfield  {journal} {\bibinfo
  {journal} {Nature Communications}\ }\textbf {\bibinfo {volume} {5}},\
  \bibinfo {pages} {4351} (\bibinfo {year} {2014})}\BibitemShut {NoStop}%
\bibitem [{\citenamefont {Weber}\ \emph {et~al.}(2018)\citenamefont {Weber},
  \citenamefont {Rycroft},\ and\ \citenamefont {Mahadevan}}]{Weber2018}%
  \BibitemOpen
  \bibfield  {author} {\bibinfo {author} {\bibfnamefont {C.~A.}\ \bibnamefont
  {Weber}}, \bibinfo {author} {\bibfnamefont {C.~H.}\ \bibnamefont {Rycroft}},\
  and\ \bibinfo {author} {\bibfnamefont {L.}~\bibnamefont {Mahadevan}},\ }\href
  {https://doi.org/10.1103/PhysRevLett.120.248003} {\bibfield  {journal}
  {\bibinfo  {journal} {Physical Review Letters}\ }\textbf {\bibinfo {volume}
  {120}},\ \bibinfo {pages} {248003} (\bibinfo {year} {2018})}\BibitemShut
  {NoStop}%
\bibitem [{\citenamefont {Tiribocchi}\ \emph {et~al.}(2015)\citenamefont
  {Tiribocchi}, \citenamefont {Wittkowski}, \citenamefont {Marenduzzo},\ and\
  \citenamefont {Cates}}]{Tiribocchi2015}%
  \BibitemOpen
  \bibfield  {author} {\bibinfo {author} {\bibfnamefont {A.}~\bibnamefont
  {Tiribocchi}}, \bibinfo {author} {\bibfnamefont {R.}~\bibnamefont
  {Wittkowski}}, \bibinfo {author} {\bibfnamefont {D.}~\bibnamefont
  {Marenduzzo}},\ and\ \bibinfo {author} {\bibfnamefont {M.~E.}\ \bibnamefont
  {Cates}},\ }\href {https://doi.org/10.1103/PhysRevLett.115.188302} {\bibfield
   {journal} {\bibinfo  {journal} {Physical Review Letters}\ }\textbf {\bibinfo
  {volume} {115}},\ \bibinfo {pages} {188302} (\bibinfo {year}
  {2015})}\BibitemShut {NoStop}%
\bibitem [{\citenamefont {Graham}\ \emph {et~al.}(2024)\citenamefont {Graham},
  \citenamefont {Zhang},\ and\ \citenamefont {Yeomans}}]{Graham2024}%
  \BibitemOpen
  \bibfield  {author} {\bibinfo {author} {\bibfnamefont {J.~N.}\ \bibnamefont
  {Graham}}, \bibinfo {author} {\bibfnamefont {G.~M.}\ \bibnamefont {Zhang}},\
  and\ \bibinfo {author} {\bibfnamefont {J.~M.}\ \bibnamefont {Yeomans}},\
  }\href {https://doi.org/10.1039/D3SM01033C} {\bibfield  {journal} {\bibinfo
  {journal} {Soft Matter}\ }\textbf {\bibinfo {volume} {20}},\ \bibinfo {pages}
  {2955} (\bibinfo {year} {2024})}\BibitemShut {NoStop}%
\bibitem [{\citenamefont {Rozman}\ and\ \citenamefont
  {Yeomans}(2024)}]{Rozman2024}%
  \BibitemOpen
  \bibfield  {author} {\bibinfo {author} {\bibfnamefont {J.}~\bibnamefont
  {Rozman}}\ and\ \bibinfo {author} {\bibfnamefont {J.~M.}\ \bibnamefont
  {Yeomans}},\ }\href {https://doi.org/10.1103/PhysRevLett.133.248401}
  {\bibfield  {journal} {\bibinfo  {journal} {Physical Review Letters}\
  }\textbf {\bibinfo {volume} {133}},\ \bibinfo {pages} {248401} (\bibinfo
  {year} {2024})}\BibitemShut {NoStop}%
\bibitem [{\citenamefont {Yang}\ \emph {et~al.}(2014)\citenamefont {Yang},
  \citenamefont {Manning},\ and\ \citenamefont {Marchetti}}]{Yang2014}%
  \BibitemOpen
  \bibfield  {author} {\bibinfo {author} {\bibfnamefont {X.~B.}\ \bibnamefont
  {Yang}}, \bibinfo {author} {\bibfnamefont {M.~L.}\ \bibnamefont {Manning}},\
  and\ \bibinfo {author} {\bibfnamefont {M.~C.}\ \bibnamefont {Marchetti}},\
  }\href {https://doi.org/10.1039/C4SM00927D} {\bibfield  {journal} {\bibinfo
  {journal} {Soft Matter}\ }\textbf {\bibinfo {volume} {10}},\ \bibinfo {pages}
  {6477} (\bibinfo {year} {2014})}\BibitemShut {NoStop}%
\bibitem [{\citenamefont {Neville}\ \emph {et~al.}(2024)\citenamefont
  {Neville}, \citenamefont {Eggers},\ and\ \citenamefont
  {Liverpool}}]{Neville2024}%
  \BibitemOpen
  \bibfield  {author} {\bibinfo {author} {\bibfnamefont {L.}~\bibnamefont
  {Neville}}, \bibinfo {author} {\bibfnamefont {J.}~\bibnamefont {Eggers}},\
  and\ \bibinfo {author} {\bibfnamefont {T.~B.}\ \bibnamefont {Liverpool}},\
  }\href@noop {} {\bibfield  {journal} {\bibinfo  {journal} {Soft Matter}\
  }\textbf {\bibinfo {volume} {20}},\ \bibinfo {pages} {8395} (\bibinfo {year}
  {2024})}\BibitemShut {NoStop}%
\bibitem [{\citenamefont {Pérez-González}\ \emph {et~al.}(2019)\citenamefont
  {Pérez-González}, \citenamefont {Alert}, \citenamefont {Blanch-Mercader},
  \citenamefont {Gómez-González}, \citenamefont {Kolodziej}, \citenamefont
  {Bazellieres}, \citenamefont {Casademunt},\ and\ \citenamefont
  {Trepat}}]{PerezGonzalez2019}%
  \BibitemOpen
  \bibfield  {author} {\bibinfo {author} {\bibfnamefont {C.}~\bibnamefont
  {Pérez-González}}, \bibinfo {author} {\bibfnamefont {R.}~\bibnamefont
  {Alert}}, \bibinfo {author} {\bibfnamefont {C.}~\bibnamefont
  {Blanch-Mercader}}, \bibinfo {author} {\bibfnamefont {M.}~\bibnamefont
  {Gómez-González}}, \bibinfo {author} {\bibfnamefont {T.}~\bibnamefont
  {Kolodziej}}, \bibinfo {author} {\bibfnamefont {E.}~\bibnamefont
  {Bazellieres}}, \bibinfo {author} {\bibfnamefont {J.}~\bibnamefont
  {Casademunt}},\ and\ \bibinfo {author} {\bibfnamefont {X.}~\bibnamefont
  {Trepat}},\ }\href@noop {} {\bibfield  {journal} {\bibinfo  {journal} {Nature
  Physics}\ }\textbf {\bibinfo {volume} {15}},\ \bibinfo {pages} {79} (\bibinfo
  {year} {2019})}\BibitemShut {NoStop}%
\bibitem [{\citenamefont {Berke}\ \emph {et~al.}(2008)\citenamefont {Berke},
  \citenamefont {Turner}, \citenamefont {Berg},\ and\ \citenamefont
  {Lauga}}]{Berke2008}%
  \BibitemOpen
  \bibfield  {author} {\bibinfo {author} {\bibfnamefont {A.~P.}\ \bibnamefont
  {Berke}}, \bibinfo {author} {\bibfnamefont {L.}~\bibnamefont {Turner}},
  \bibinfo {author} {\bibfnamefont {H.~C.}\ \bibnamefont {Berg}},\ and\
  \bibinfo {author} {\bibfnamefont {E.}~\bibnamefont {Lauga}},\ }\href
  {https://doi.org/10.1103/PhysRevLett.101.038102} {\bibfield  {journal}
  {\bibinfo  {journal} {Physical Review Letters}\ }\textbf {\bibinfo {volume}
  {101}},\ \bibinfo {pages} {038102} (\bibinfo {year} {2008})}\BibitemShut
  {NoStop}%
\bibitem [{\citenamefont {Spagnolie}\ and\ \citenamefont
  {Lauga}(2012)}]{Spagnolie2012}%
  \BibitemOpen
  \bibfield  {author} {\bibinfo {author} {\bibfnamefont {S.~E.}\ \bibnamefont
  {Spagnolie}}\ and\ \bibinfo {author} {\bibfnamefont {E.}~\bibnamefont
  {Lauga}},\ }\href {https://doi.org/10.1017/jfm.2012.101} {\bibfield
  {journal} {\bibinfo  {journal} {Journal of Fluid Mechanics}\ }\textbf
  {\bibinfo {volume} {700}},\ \bibinfo {pages} {105–147} (\bibinfo {year}
  {2012})}\BibitemShut {NoStop}%
\bibitem [{\citenamefont {Doostmohammadi}\ \emph {et~al.}(2018)\citenamefont
  {Doostmohammadi}, \citenamefont {Ign{\'e}s-Mullol}, \citenamefont {Yeomans},\
  and\ \citenamefont {Sagu{\'e}s}}]{doostmohammadi2018active}%
  \BibitemOpen
  \bibfield  {author} {\bibinfo {author} {\bibfnamefont {A.}~\bibnamefont
  {Doostmohammadi}}, \bibinfo {author} {\bibfnamefont {J.}~\bibnamefont
  {Ign{\'e}s-Mullol}}, \bibinfo {author} {\bibfnamefont {J.~M.}\ \bibnamefont
  {Yeomans}},\ and\ \bibinfo {author} {\bibfnamefont {F.}~\bibnamefont
  {Sagu{\'e}s}},\ }\href@noop {} {\bibfield  {journal} {\bibinfo  {journal}
  {Nature Communications}\ }\textbf {\bibinfo {volume} {9}},\ \bibinfo {pages}
  {3246} (\bibinfo {year} {2018})}\BibitemShut {NoStop}%
\bibitem [{\citenamefont {Saw}\ \emph {et~al.}(2017)\citenamefont {Saw},
  \citenamefont {Doostmohammadi}, \citenamefont {Nier}, \citenamefont
  {Kocgozlu}, \citenamefont {Thampi}, \citenamefont {Toyama}, \citenamefont
  {Marcq}, \citenamefont {Lim}, \citenamefont {Yeomans},\ and\ \citenamefont
  {Ladoux}}]{Saw2017}%
  \BibitemOpen
  \bibfield  {author} {\bibinfo {author} {\bibfnamefont {T.~B.}\ \bibnamefont
  {Saw}}, \bibinfo {author} {\bibfnamefont {A.}~\bibnamefont {Doostmohammadi}},
  \bibinfo {author} {\bibfnamefont {V.}~\bibnamefont {Nier}}, \bibinfo {author}
  {\bibfnamefont {L.}~\bibnamefont {Kocgozlu}}, \bibinfo {author}
  {\bibfnamefont {S.}~\bibnamefont {Thampi}}, \bibinfo {author} {\bibfnamefont
  {Y.}~\bibnamefont {Toyama}}, \bibinfo {author} {\bibfnamefont
  {P.}~\bibnamefont {Marcq}}, \bibinfo {author} {\bibfnamefont {C.~T.}\
  \bibnamefont {Lim}}, \bibinfo {author} {\bibfnamefont {J.~M.}\ \bibnamefont
  {Yeomans}},\ and\ \bibinfo {author} {\bibfnamefont {B.}~\bibnamefont
  {Ladoux}},\ }\href {https://doi.org/10.1038/nature21718} {\bibfield
  {journal} {\bibinfo  {journal} {Nature}\ }\textbf {\bibinfo {volume} {544}},\
  \bibinfo {pages} {212} (\bibinfo {year} {2017})}\BibitemShut {NoStop}%
\bibitem [{\citenamefont {Duclos}\ \emph {et~al.}(2018)\citenamefont {Duclos},
  \citenamefont {Blanch-Mercader}, \citenamefont {Yashunsky}, \citenamefont
  {Salbreux}, \citenamefont {Joanny}, \citenamefont {Prost},\ and\
  \citenamefont {Silberzan}}]{duclos2018spontaneous}%
  \BibitemOpen
  \bibfield  {author} {\bibinfo {author} {\bibfnamefont {G.}~\bibnamefont
  {Duclos}}, \bibinfo {author} {\bibfnamefont {C.}~\bibnamefont
  {Blanch-Mercader}}, \bibinfo {author} {\bibfnamefont {V.}~\bibnamefont
  {Yashunsky}}, \bibinfo {author} {\bibfnamefont {G.}~\bibnamefont {Salbreux}},
  \bibinfo {author} {\bibfnamefont {J.-F.}\ \bibnamefont {Joanny}}, \bibinfo
  {author} {\bibfnamefont {J.}~\bibnamefont {Prost}},\ and\ \bibinfo {author}
  {\bibfnamefont {P.}~\bibnamefont {Silberzan}},\ }\href@noop {} {\bibfield
  {journal} {\bibinfo  {journal} {Nature Physics}\ }\textbf {\bibinfo {volume}
  {14}},\ \bibinfo {pages} {728} (\bibinfo {year} {2018})}\BibitemShut
  {NoStop}%
\bibitem [{\citenamefont {Simha}\ and\ \citenamefont
  {Ramaswamy}(2002)}]{SimhaRamaswamy2002}%
  \BibitemOpen
  \bibfield  {author} {\bibinfo {author} {\bibfnamefont {R.~A.}\ \bibnamefont
  {Simha}}\ and\ \bibinfo {author} {\bibfnamefont {S.}~\bibnamefont
  {Ramaswamy}},\ }\href {https://doi.org/10.1103/PhysRevLett.89.058101}
  {\bibfield  {journal} {\bibinfo  {journal} {Physical Review Letters}\
  }\textbf {\bibinfo {volume} {89}},\ \bibinfo {pages} {058101} (\bibinfo
  {year} {2002})}\BibitemShut {NoStop}%
\bibitem [{\citenamefont {Voituriez}\ \emph {et~al.}(2005)\citenamefont
  {Voituriez}, \citenamefont {Joanny},\ and\ \citenamefont
  {Prost}}]{voituriez2005spontaneous}%
  \BibitemOpen
  \bibfield  {author} {\bibinfo {author} {\bibfnamefont {R.}~\bibnamefont
  {Voituriez}}, \bibinfo {author} {\bibfnamefont {J.-F.}\ \bibnamefont
  {Joanny}},\ and\ \bibinfo {author} {\bibfnamefont {J.}~\bibnamefont
  {Prost}},\ }\href@noop {} {\bibfield  {journal} {\bibinfo  {journal}
  {Europhysics Letters}\ }\textbf {\bibinfo {volume} {70}},\ \bibinfo {pages}
  {404} (\bibinfo {year} {2005})}\BibitemShut {NoStop}%
\bibitem [{\citenamefont {Shendruk}\ \emph {et~al.}(2017)\citenamefont
  {Shendruk}, \citenamefont {Doostmohammadi}, \citenamefont {Thijssen},\ and\
  \citenamefont {Yeomans}}]{shendruk2017dancing}%
  \BibitemOpen
  \bibfield  {author} {\bibinfo {author} {\bibfnamefont {T.~N.}\ \bibnamefont
  {Shendruk}}, \bibinfo {author} {\bibfnamefont {A.}~\bibnamefont
  {Doostmohammadi}}, \bibinfo {author} {\bibfnamefont {K.}~\bibnamefont
  {Thijssen}},\ and\ \bibinfo {author} {\bibfnamefont {J.~M.}\ \bibnamefont
  {Yeomans}},\ }\href@noop {} {\bibfield  {journal} {\bibinfo  {journal} {Soft
  Matter}\ }\textbf {\bibinfo {volume} {13}},\ \bibinfo {pages} {3853}
  (\bibinfo {year} {2017})}\BibitemShut {NoStop}%
\bibitem [{\citenamefont {Opathalage}\ \emph {et~al.}(2019)\citenamefont
  {Opathalage}, \citenamefont {Norton}, \citenamefont {Juniper}, \citenamefont
  {Langeslay}, \citenamefont {Aghvami}, \citenamefont {Fraden},\ and\
  \citenamefont {Dogic}}]{opathalage2019self}%
  \BibitemOpen
  \bibfield  {author} {\bibinfo {author} {\bibfnamefont {A.}~\bibnamefont
  {Opathalage}}, \bibinfo {author} {\bibfnamefont {M.~M.}\ \bibnamefont
  {Norton}}, \bibinfo {author} {\bibfnamefont {M.~P.~N.}\ \bibnamefont
  {Juniper}}, \bibinfo {author} {\bibfnamefont {B.}~\bibnamefont {Langeslay}},
  \bibinfo {author} {\bibfnamefont {S.~A.}\ \bibnamefont {Aghvami}}, \bibinfo
  {author} {\bibfnamefont {S.}~\bibnamefont {Fraden}},\ and\ \bibinfo {author}
  {\bibfnamefont {Z.}~\bibnamefont {Dogic}},\ }\href@noop {} {\bibfield
  {journal} {\bibinfo  {journal} {Proceedings of the National Academy of
  Sciences of the United States of America}\ }\textbf {\bibinfo {volume}
  {116}},\ \bibinfo {pages} {4788} (\bibinfo {year} {2019})}\BibitemShut
  {NoStop}%
\bibitem [{\citenamefont {Caballero}\ \emph {et~al.}(2023)\citenamefont
  {Caballero}, \citenamefont {You},\ and\ \citenamefont
  {Marchetti}}]{Caballero2023}%
  \BibitemOpen
  \bibfield  {author} {\bibinfo {author} {\bibfnamefont {F.}~\bibnamefont
  {Caballero}}, \bibinfo {author} {\bibfnamefont {Z.}~\bibnamefont {You}},\
  and\ \bibinfo {author} {\bibfnamefont {M.~C.}\ \bibnamefont {Marchetti}},\
  }\href {https://doi.org/10.1039/d3sm00744h} {\bibfield  {journal} {\bibinfo
  {journal} {Soft Matter}\ }\textbf {\bibinfo {volume} {19}},\ \bibinfo {pages}
  {7828–7835} (\bibinfo {year} {2023})}\BibitemShut {NoStop}%
\bibitem [{\citenamefont {Blow}\ \emph {et~al.}(2014)\citenamefont {Blow},
  \citenamefont {Thampi},\ and\ \citenamefont {Yeomans}}]{Blow2014}%
  \BibitemOpen
  \bibfield  {author} {\bibinfo {author} {\bibfnamefont {M.~L.}\ \bibnamefont
  {Blow}}, \bibinfo {author} {\bibfnamefont {S.~P.}\ \bibnamefont {Thampi}},\
  and\ \bibinfo {author} {\bibfnamefont {J.~M.}\ \bibnamefont {Yeomans}},\
  }\href {https://doi.org/10.1103/PhysRevLett.113.248303} {\bibfield  {journal}
  {\bibinfo  {journal} {Physical Review Letters}\ }\textbf {\bibinfo {volume}
  {113}},\ \bibinfo {pages} {248303} (\bibinfo {year} {2014})}\BibitemShut
  {NoStop}%
\bibitem [{\citenamefont {Soni}\ \emph {et~al.}(2019)\citenamefont {Soni},
  \citenamefont {Luo}, \citenamefont {Pelcovits},\ and\ \citenamefont
  {Powers}}]{Soni2019}%
  \BibitemOpen
  \bibfield  {author} {\bibinfo {author} {\bibfnamefont {H.}~\bibnamefont
  {Soni}}, \bibinfo {author} {\bibfnamefont {W.}~\bibnamefont {Luo}}, \bibinfo
  {author} {\bibfnamefont {R.~A.}\ \bibnamefont {Pelcovits}},\ and\ \bibinfo
  {author} {\bibfnamefont {T.~R.}\ \bibnamefont {Powers}},\ }\href
  {https://doi.org/10.1039/c9sm01216h} {\bibfield  {journal} {\bibinfo
  {journal} {Soft Matter}\ }\textbf {\bibinfo {volume} {15}},\ \bibinfo {pages}
  {6318–6330} (\bibinfo {year} {2019})}\BibitemShut {NoStop}%
\bibitem [{\citenamefont {Gulati}\ \emph {et~al.}(2024)\citenamefont {Gulati},
  \citenamefont {Caballero}, \citenamefont {Kolvin}, \citenamefont {You},\ and\
  \citenamefont {Marchetti}}]{Gulati2024}%
  \BibitemOpen
  \bibfield  {author} {\bibinfo {author} {\bibfnamefont {P.}~\bibnamefont
  {Gulati}}, \bibinfo {author} {\bibfnamefont {F.}~\bibnamefont {Caballero}},
  \bibinfo {author} {\bibfnamefont {I.}~\bibnamefont {Kolvin}}, \bibinfo
  {author} {\bibfnamefont {Z.}~\bibnamefont {You}},\ and\ \bibinfo {author}
  {\bibfnamefont {M.~C.}\ \bibnamefont {Marchetti}},\ }\href
  {https://doi.org/10.1039/d4sm00822g} {\bibfield  {journal} {\bibinfo
  {journal} {Soft Matter}\ }\textbf {\bibinfo {volume} {20}},\ \bibinfo {pages}
  {7703–7714} (\bibinfo {year} {2024})}\BibitemShut {NoStop}%
\bibitem [{\citenamefont {Adkins}\ \emph {et~al.}(2022)\citenamefont {Adkins},
  \citenamefont {Kolvin}, \citenamefont {You}, \citenamefont {Witthaus},
  \citenamefont {Marchetti},\ and\ \citenamefont {Dogic}}]{Adkins2022}%
  \BibitemOpen
  \bibfield  {author} {\bibinfo {author} {\bibfnamefont {R.}~\bibnamefont
  {Adkins}}, \bibinfo {author} {\bibfnamefont {I.}~\bibnamefont {Kolvin}},
  \bibinfo {author} {\bibfnamefont {Z.~H.}\ \bibnamefont {You}}, \bibinfo
  {author} {\bibfnamefont {S.}~\bibnamefont {Witthaus}}, \bibinfo {author}
  {\bibfnamefont {M.~C.}\ \bibnamefont {Marchetti}},\ and\ \bibinfo {author}
  {\bibfnamefont {Z.}~\bibnamefont {Dogic}},\ }\href@noop {} {\bibfield
  {journal} {\bibinfo  {journal} {Science}\ }\textbf {\bibinfo {volume}
  {377}},\ \bibinfo {pages} {768} (\bibinfo {year} {2022})}\BibitemShut
  {NoStop}%
\bibitem [{\citenamefont {Alert}\ \emph {et~al.}(2019)\citenamefont {Alert},
  \citenamefont {Blanch-Mercader},\ and\ \citenamefont
  {Casademunt}}]{Alert2019}%
  \BibitemOpen
  \bibfield  {author} {\bibinfo {author} {\bibfnamefont {R.}~\bibnamefont
  {Alert}}, \bibinfo {author} {\bibfnamefont {C.}~\bibnamefont
  {Blanch-Mercader}},\ and\ \bibinfo {author} {\bibfnamefont {J.}~\bibnamefont
  {Casademunt}},\ }\href {https://doi.org/10.1103/PhysRevLett.122.088104}
  {\bibfield  {journal} {\bibinfo  {journal} {Physical Review Letters}\
  }\textbf {\bibinfo {volume} {122}},\ \bibinfo {pages} {088104} (\bibinfo
  {year} {2019})}\BibitemShut {NoStop}%
\bibitem [{\citenamefont {Zhao}\ \emph {et~al.}(2024)\citenamefont {Zhao},
  \citenamefont {Gulati}, \citenamefont {Caballero}, \citenamefont {Kolvin},
  \citenamefont {Adkins}, \citenamefont {Marchetti},\ and\ \citenamefont
  {Dogic}}]{zhao2024asymmetric}%
  \BibitemOpen
  \bibfield  {author} {\bibinfo {author} {\bibfnamefont {L.}~\bibnamefont
  {Zhao}}, \bibinfo {author} {\bibfnamefont {P.}~\bibnamefont {Gulati}},
  \bibinfo {author} {\bibfnamefont {F.}~\bibnamefont {Caballero}}, \bibinfo
  {author} {\bibfnamefont {I.}~\bibnamefont {Kolvin}}, \bibinfo {author}
  {\bibfnamefont {R.}~\bibnamefont {Adkins}}, \bibinfo {author} {\bibfnamefont
  {M.~C.}\ \bibnamefont {Marchetti}},\ and\ \bibinfo {author} {\bibfnamefont
  {Z.}~\bibnamefont {Dogic}},\ }\href@noop {} {\bibfield  {journal} {\bibinfo
  {journal} {Proceedings of the National Academy of Sciences of the United
  States of America}\ }\textbf {\bibinfo {volume} {121}},\ \bibinfo {pages}
  {2410345121} (\bibinfo {year} {2024})}\BibitemShut {NoStop}%
\bibitem [{\citenamefont {Bhattacharyya}\ and\ \citenamefont
  {Yeomans}(2023)}]{Bhattacharyya_2023}%
  \BibitemOpen
  \bibfield  {author} {\bibinfo {author} {\bibfnamefont {S.}~\bibnamefont
  {Bhattacharyya}}\ and\ \bibinfo {author} {\bibfnamefont {J.~M.}\ \bibnamefont
  {Yeomans}},\ }\href {https://doi.org/10.1103/physrevlett.130.238201}
  {\bibfield  {journal} {\bibinfo  {journal} {Physical Review Letters}\
  }\textbf {\bibinfo {volume} {130}},\ \bibinfo {pages} {238201} (\bibinfo
  {year} {2023})}\BibitemShut {NoStop}%
\bibitem [{\citenamefont {Joanny}\ \emph {et~al.}(2007)\citenamefont {Joanny},
  \citenamefont {Jülicher}, \citenamefont {Kruse},\ and\ \citenamefont
  {Prost}}]{Joanny2007}%
  \BibitemOpen
  \bibfield  {author} {\bibinfo {author} {\bibfnamefont {J.~F.}\ \bibnamefont
  {Joanny}}, \bibinfo {author} {\bibfnamefont {F.}~\bibnamefont {Jülicher}},
  \bibinfo {author} {\bibfnamefont {K.}~\bibnamefont {Kruse}},\ and\ \bibinfo
  {author} {\bibfnamefont {J.}~\bibnamefont {Prost}},\ }\href
  {https://doi.org/10.1088/1367-2630/9/11/422} {\bibfield  {journal} {\bibinfo
  {journal} {New Journal of Physics}\ }\textbf {\bibinfo {volume} {9}},\
  \bibinfo {pages} {422} (\bibinfo {year} {2007})}\BibitemShut {NoStop}%
\bibitem [{\citenamefont {{S. Bhattacharyya and J. M.
  Yeomans}}(2024)}]{bhattacharyya2024phase}%
  \BibitemOpen
  \bibfield  {author} {\bibinfo {author} {\bibnamefont {{S. Bhattacharyya and
  J. M. Yeomans}}},\ }\href {https://doi.org/10.1103/PhysRevE.110.024607}
  {\bibfield  {journal} {\bibinfo  {journal} {Physical Review E}\ }\textbf
  {\bibinfo {volume} {110}},\ \bibinfo {pages} {024607} (\bibinfo {year}
  {2024})}\BibitemShut {NoStop}%
\bibitem [{\citenamefont {Malevanets}\ and\ \citenamefont
  {M.~Yeomans}(1999)}]{Malevanets1999}%
  \BibitemOpen
  \bibfield  {author} {\bibinfo {author} {\bibfnamefont {A.}~\bibnamefont
  {Malevanets}}\ and\ \bibinfo {author} {\bibfnamefont {J.}~\bibnamefont
  {M.~Yeomans}},\ }\href {https://doi.org/10.1039/A809152H} {\bibfield
  {journal} {\bibinfo  {journal} {Faraday Discussions}\ }\textbf {\bibinfo
  {volume} {112}},\ \bibinfo {pages} {237} (\bibinfo {year}
  {1999})}\BibitemShut {NoStop}%
\bibitem [{\citenamefont {Malevanets}\ and\ \citenamefont
  {M.~Yeomans}(2000)}]{Malevanets2000}%
  \BibitemOpen
  \bibfield  {author} {\bibinfo {author} {\bibfnamefont {A.}~\bibnamefont
  {Malevanets}}\ and\ \bibinfo {author} {\bibfnamefont {J.}~\bibnamefont
  {M.~Yeomans}},\ }\href
  {https://doi.org/https://doi.org/10.1016/S0010-4655(00)00030-8} {\bibfield
  {journal} {\bibinfo  {journal} {Computer Physics Communications}\ }\textbf
  {\bibinfo {volume} {127}},\ \bibinfo {pages} {105} (\bibinfo {year}
  {2000})}\BibitemShut {NoStop}%
\bibitem [{\citenamefont {De~Gennes}\ and\ \citenamefont
  {Prost}(1993)}]{degennes_book}%
  \BibitemOpen
  \bibfield  {author} {\bibinfo {author} {\bibfnamefont {P.~G.}\ \bibnamefont
  {De~Gennes}}\ and\ \bibinfo {author} {\bibfnamefont {J.}~\bibnamefont
  {Prost}},\ }\href {https://books.google.co.uk/books?id=0Nw-dzWz5agC} {\emph
  {\bibinfo {title} {The Physics of Liquid Crystals}}}\ (\bibinfo  {publisher}
  {Oxford University Press},\ \bibinfo {year} {1993})\BibitemShut {NoStop}%
\bibitem [{\citenamefont {Beris}\ and\ \citenamefont
  {Edwards}(1994)}]{beris1994thermodynamics}%
  \BibitemOpen
  \bibfield  {author} {\bibinfo {author} {\bibfnamefont {A.~N.}\ \bibnamefont
  {Beris}}\ and\ \bibinfo {author} {\bibfnamefont {B.~J.}\ \bibnamefont
  {Edwards}},\ }\href@noop {} {\emph {\bibinfo {title} {Thermodynamics of
  flowing systems: with internal microstructure}}}\ (\bibinfo  {publisher}
  {Oxford University Press},\ \bibinfo {year} {1994})\BibitemShut {NoStop}%
\bibitem [{\citenamefont {Pokawanvit}\ \emph {et~al.}(2022)\citenamefont
  {Pokawanvit}, \citenamefont {Chen}, \citenamefont {You}, \citenamefont
  {Angheluta}, \citenamefont {Marchetti},\ and\ \citenamefont
  {Bowick}}]{Supavit2022}%
  \BibitemOpen
  \bibfield  {author} {\bibinfo {author} {\bibfnamefont {S.}~\bibnamefont
  {Pokawanvit}}, \bibinfo {author} {\bibfnamefont {Z.}~\bibnamefont {Chen}},
  \bibinfo {author} {\bibfnamefont {Z.}~\bibnamefont {You}}, \bibinfo {author}
  {\bibfnamefont {L.}~\bibnamefont {Angheluta}}, \bibinfo {author}
  {\bibfnamefont {M.~C.}\ \bibnamefont {Marchetti}},\ and\ \bibinfo {author}
  {\bibfnamefont {M.~J.}\ \bibnamefont {Bowick}},\ }\href
  {https://doi.org/10.1103/PhysRevE.106.054610} {\bibfield  {journal} {\bibinfo
   {journal} {Physical Review E}\ }\textbf {\bibinfo {volume} {106}},\ \bibinfo
  {pages} {054610} (\bibinfo {year} {2022})}\BibitemShut {NoStop}%
\bibitem [{\citenamefont {Krueger}\ \emph {et~al.}(2016)\citenamefont
  {Krueger}, \citenamefont {Kusumaatmaja}, \citenamefont {Kuzmin},
  \citenamefont {Shardt}, \citenamefont {Silva},\ and\ \citenamefont
  {Viggen}}]{LB_book}%
  \BibitemOpen
  \bibfield  {author} {\bibinfo {author} {\bibfnamefont {T.}~\bibnamefont
  {Krueger}}, \bibinfo {author} {\bibfnamefont {H.}~\bibnamefont
  {Kusumaatmaja}}, \bibinfo {author} {\bibfnamefont {A.}~\bibnamefont
  {Kuzmin}}, \bibinfo {author} {\bibfnamefont {O.}~\bibnamefont {Shardt}},
  \bibinfo {author} {\bibfnamefont {G.}~\bibnamefont {Silva}},\ and\ \bibinfo
  {author} {\bibfnamefont {E.}~\bibnamefont {Viggen}},\ }\href@noop {} {\emph
  {\bibinfo {title} {The Lattice Boltzmann Method: Principles and Practice}}},\
  Graduate Texts in Physics\ (\bibinfo  {publisher} {Springer},\ \bibinfo
  {year} {2016})\BibitemShut {NoStop}%
\bibitem [{\citenamefont {Coelho}\ \emph {et~al.}(2023)\citenamefont {Coelho},
  \citenamefont {Figueiredo},\ and\ \citenamefont {Telo~da Gama}}]{Coelho2023}%
  \BibitemOpen
  \bibfield  {author} {\bibinfo {author} {\bibfnamefont {R.~C.~V.}\
  \bibnamefont {Coelho}}, \bibinfo {author} {\bibfnamefont {H.~R. J.~C.}\
  \bibnamefont {Figueiredo}},\ and\ \bibinfo {author} {\bibfnamefont {M.~M.}\
  \bibnamefont {Telo~da Gama}},\ }\href
  {https://doi.org/10.1103/PhysRevResearch.5.033165} {\bibfield  {journal}
  {\bibinfo  {journal} {Physical Review Research}\ }\textbf {\bibinfo {volume}
  {5}},\ \bibinfo {pages} {033165} (\bibinfo {year} {2023})}\BibitemShut
  {NoStop}%
\bibitem [{\citenamefont {Coelho}\ \emph {et~al.}(2021)\citenamefont {Coelho},
  \citenamefont {Araújo},\ and\ \citenamefont {Telo~da Gama}}]{Coelho2021}%
  \BibitemOpen
  \bibfield  {author} {\bibinfo {author} {\bibfnamefont {R.~C.~V.}\
  \bibnamefont {Coelho}}, \bibinfo {author} {\bibfnamefont {N.~A.~M.}\
  \bibnamefont {Araújo}},\ and\ \bibinfo {author} {\bibfnamefont {M.~M.}\
  \bibnamefont {Telo~da Gama}},\ }\href@noop {} {\bibfield  {journal} {\bibinfo
   {journal} {Philosophical Transactions of the Royal Society A: Mathematical,
  Physical and Engineering Sciences}\ }\textbf {\bibinfo {volume} {379}},\
  \bibinfo {pages} {20200394} (\bibinfo {year} {2021})}\BibitemShut {NoStop}%
\bibitem [{\citenamefont {Santhosh}\ \emph {et~al.}(2020)\citenamefont
  {Santhosh}, \citenamefont {Nejad}, \citenamefont {Doostmohammadi},
  \citenamefont {Yeomans},\ and\ \citenamefont {Thampi}}]{Santhosh2020}%
  \BibitemOpen
  \bibfield  {author} {\bibinfo {author} {\bibfnamefont {S.}~\bibnamefont
  {Santhosh}}, \bibinfo {author} {\bibfnamefont {M.~R.}\ \bibnamefont {Nejad}},
  \bibinfo {author} {\bibfnamefont {A.}~\bibnamefont {Doostmohammadi}},
  \bibinfo {author} {\bibfnamefont {J.~M.}\ \bibnamefont {Yeomans}},\ and\
  \bibinfo {author} {\bibfnamefont {S.~P.}\ \bibnamefont {Thampi}},\ }\href
  {https://doi.org/10.1007/s10955-020-02497-0} {\bibfield  {journal} {\bibinfo
  {journal} {Journal of Statistical Physics}\ }\textbf {\bibinfo {volume}
  {180}},\ \bibinfo {pages} {699} (\bibinfo {year} {2020})}\BibitemShut
  {NoStop}%
\bibitem [{\citenamefont {Woodhouse}\ and\ \citenamefont
  {Goldstein}(2012)}]{Woodhouse2012}%
  \BibitemOpen
  \bibfield  {author} {\bibinfo {author} {\bibfnamefont {F.~G.}\ \bibnamefont
  {Woodhouse}}\ and\ \bibinfo {author} {\bibfnamefont {R.~E.}\ \bibnamefont
  {Goldstein}},\ }\href@noop {} {\bibfield  {journal} {\bibinfo  {journal}
  {Physical Review Letters}\ }\textbf {\bibinfo {volume} {109}},\ \bibinfo
  {pages} {168105} (\bibinfo {year} {2012})}\BibitemShut {NoStop}%
\bibitem [{\citenamefont {Townes}\ and\ \citenamefont
  {Holtfreter}(1955)}]{Townes1955amphibian}%
  \BibitemOpen
  \bibfield  {author} {\bibinfo {author} {\bibfnamefont {P.~L.}\ \bibnamefont
  {Townes}}\ and\ \bibinfo {author} {\bibfnamefont {J.}~\bibnamefont
  {Holtfreter}},\ }\href
  {https://doi.org/https://doi.org/10.1002/jez.1401280105} {\bibfield
  {journal} {\bibinfo  {journal} {Journal of Experimental Zoology}\ }\textbf
  {\bibinfo {volume} {128}},\ \bibinfo {pages} {53} (\bibinfo {year}
  {1955})}\BibitemShut {NoStop}%
\bibitem [{\citenamefont {Moore}\ \emph {et~al.}(2009)\citenamefont {Moore},
  \citenamefont {Cai}, \citenamefont {Escudero},\ and\ \citenamefont
  {Xu}}]{Moore2009}%
  \BibitemOpen
  \bibfield  {author} {\bibinfo {author} {\bibfnamefont {R.}~\bibnamefont
  {Moore}}, \bibinfo {author} {\bibfnamefont {K.~Q.}\ \bibnamefont {Cai}},
  \bibinfo {author} {\bibfnamefont {D.~O.}\ \bibnamefont {Escudero}},\ and\
  \bibinfo {author} {\bibfnamefont {X.-X.}\ \bibnamefont {Xu}},\ }\href
  {https://doi.org/https://doi.org/10.1002/dvg.20536} {\bibfield  {journal}
  {\bibinfo  {journal} {genesis}\ }\textbf {\bibinfo {volume} {47}},\ \bibinfo
  {pages} {579} (\bibinfo {year} {2009})},\ \Eprint
  {https://arxiv.org/abs/https://onlinelibrary.wiley.com/doi/pdf/10.1002/dvg.20536}
  {https://onlinelibrary.wiley.com/doi/pdf/10.1002/dvg.20536} \BibitemShut
  {NoStop}%
\bibitem [{\citenamefont {Cochet-Escartin}\ \emph {et~al.}(2017)\citenamefont
  {Cochet-Escartin}, \citenamefont {Locke}, \citenamefont {Shi}, \citenamefont
  {Steele},\ and\ \citenamefont {Collins}}]{cochetEscartin2017hydra}%
  \BibitemOpen
  \bibfield  {author} {\bibinfo {author} {\bibfnamefont {O.}~\bibnamefont
  {Cochet-Escartin}}, \bibinfo {author} {\bibfnamefont {T.~T.}\ \bibnamefont
  {Locke}}, \bibinfo {author} {\bibfnamefont {W.~H.}\ \bibnamefont {Shi}},
  \bibinfo {author} {\bibfnamefont {R.~E.}\ \bibnamefont {Steele}},\ and\
  \bibinfo {author} {\bibfnamefont {E.-M.~S.}\ \bibnamefont {Collins}},\ }\href
  {https://doi.org/10.1016/j.bpj.2017.10.045} {\bibfield  {journal} {\bibinfo
  {journal} {Biophysical Journal}\ }\textbf {\bibinfo {volume} {113}},\
  \bibinfo {pages} {2827} (\bibinfo {year} {2017})}\BibitemShut {NoStop}%
\bibitem [{\citenamefont {Batlle}\ and\ \citenamefont
  {Wilkinson}(2012)}]{batlle2012molecular}%
  \BibitemOpen
  \bibfield  {author} {\bibinfo {author} {\bibfnamefont {E.}~\bibnamefont
  {Batlle}}\ and\ \bibinfo {author} {\bibfnamefont {D.~G.}\ \bibnamefont
  {Wilkinson}},\ }\href@noop {} {\bibfield  {journal} {\bibinfo  {journal}
  {Cold Spring Harbor Perspectives in Biology}\ }\textbf {\bibinfo {volume}
  {4}},\ \bibinfo {pages} {a008227} (\bibinfo {year} {2012})}\BibitemShut
  {NoStop}%
\end{thebibliography}%

%%%%%%%%%% Merge with supplemental materials %%%%%%%%%%
\newpage
\widetext
\begin{center}
\textbf{\large Supplemental Notes:}
\end{center}
%%%%%%%%%% Merge with supplemental materials %%%%%%%%%%
%%%%%%%%%% Prefix a "S" to all equations, figures, tables and reset the counter %%%%%%%%%%
\setcounter{equation}{0}
\setcounter{figure}{0}
\setcounter{table}{0}
\setcounter{section}{0}
\makeatletter
\renewcommand{\theequation}{A\arabic{equation}}
\renewcommand{\thefigure}{A\arabic{figure}}
\renewcommand{\thesection}{A\arabic{section}}
\renewcommand{\bibnumfmt}[1]{[S#1]}
\renewcommand{\citenumfont}[1]{S#1}
%%%%%%%%%% Prefix a "S" to all equations, figures, tables and reset the counter %%%%%%%%%%

\section{Movie captions}

The eight videos illustrate the results of our simulations, and are described below. \\
 
{\bf Movie 1:} Simulations of an extensile active nematic fluid mixed with a passive isotropic fluid, confined to a square box (L=120) with imposed planar anchoring. The active fluid tends to accumulate at the boundaries.  The colourbar shows the concentration of the active component, $\phi$, black lines show the nematic orientation field and red arrows show the velocity field. Parameters take the default values listed in the text; activity is $\zeta=0.01$.\\

{\bf Movie 2:} Simulations of an extensile active nematic fluid mixed with a passive isotropic fluid, confined to a square box (L=120) with imposed homeotropic anchoring. The active fluid concentration tends decrease at the boundaries.  The colourbar shows the concentration of the active component, $\phi$, black lines show the nematic orientation field and red arrows show the velocity field. Parameters take the default values listed in the text; activity is $\zeta=0.01$.\\

{\bf Movie 3:} Simulations of an extensile active nematic fluid mixed with a passive isotropic fluid, confined to a square box (L=160). The low activity, $\zeta=0.004$ leads to weak spontaneous planar anchoring, and a small accumulation of active material at the boundary. The colourbar shows the concentration of the active component, $\phi$, black lines show the nematic orientation field and red arrows show the velocity field. Parameters take the default values listed in the text.\\

{\bf Movie 4:} Simulations of an extensile active nematic fluid mixed with a passive isotropic fluid, confined to a square box (L=160). The high activity, $\zeta=0.01$ leads to spontaneous planar anchoring, and an accumulation of active material at the boundary. The colourbar shows the concentration of the active component, $\phi$, black lines show the nematic orientation field and red arrows show the velocity field. Parameters take the default values listed in the text.\\

{\bf Movie 5:} Simulations of an extensile active nematic fluid mixed with a passive isotropic fluid, confined to a circle of radius $R=30$. Steady circulating flows are set up, and there is weak sorting of the active component to the boundary. The colourbar shows the concentration of the active component, $\phi$, black lines show the nematic orientation field and red arrows show the velocity field. Parameters take the default values listed in the text; activity is $\zeta=0.005$.\\

{\bf Movie 6:} Simulations of an extensile active nematic fluid mixed with a passive isotropic fluid, confined to a circle of radius $R=60$. Gradients in the magnitude and direction of the nematic ordering drive the active component to the boundary. The colourbar shows the concentration of the active component, $\phi$, black lines show the nematic orientation field and red arrows show the velocity field. Parameters take the default values listed in the text; activity is $\zeta=0.005$.\\

{\bf Movie 7:} Simulations of an extensile active nematic fluid mixed with a passive isotropic fluid, confined to a circle of radius $R=90$.  Gradients in the magnitude and direction of the nematic ordering drive the active component to the boundary. The colourbar shows the concentration of the active component, $\phi$, black lines show the nematic orientation field and red arrows show the velocity field. Parameters take the default values listed in the text; activity is $\zeta=0.005$.\\

{\bf Movie 8:} Simulations of an active fluid sorting to a boundary. Initially the active component is in the centre of the confining circle, stabilised by a free energy that favours thermodynamic phase ordering. When activity is turned on, the active region elongates into an elliptical shape and, on contact with the container boundary,  re-organises into a boundary ring. The colourbar shows the concentration of the active component.\\

\newpage

\section{Supplementary Figures}

\begin{figure}[htp]
    \centering
    \includegraphics[width = 0.75\textwidth]{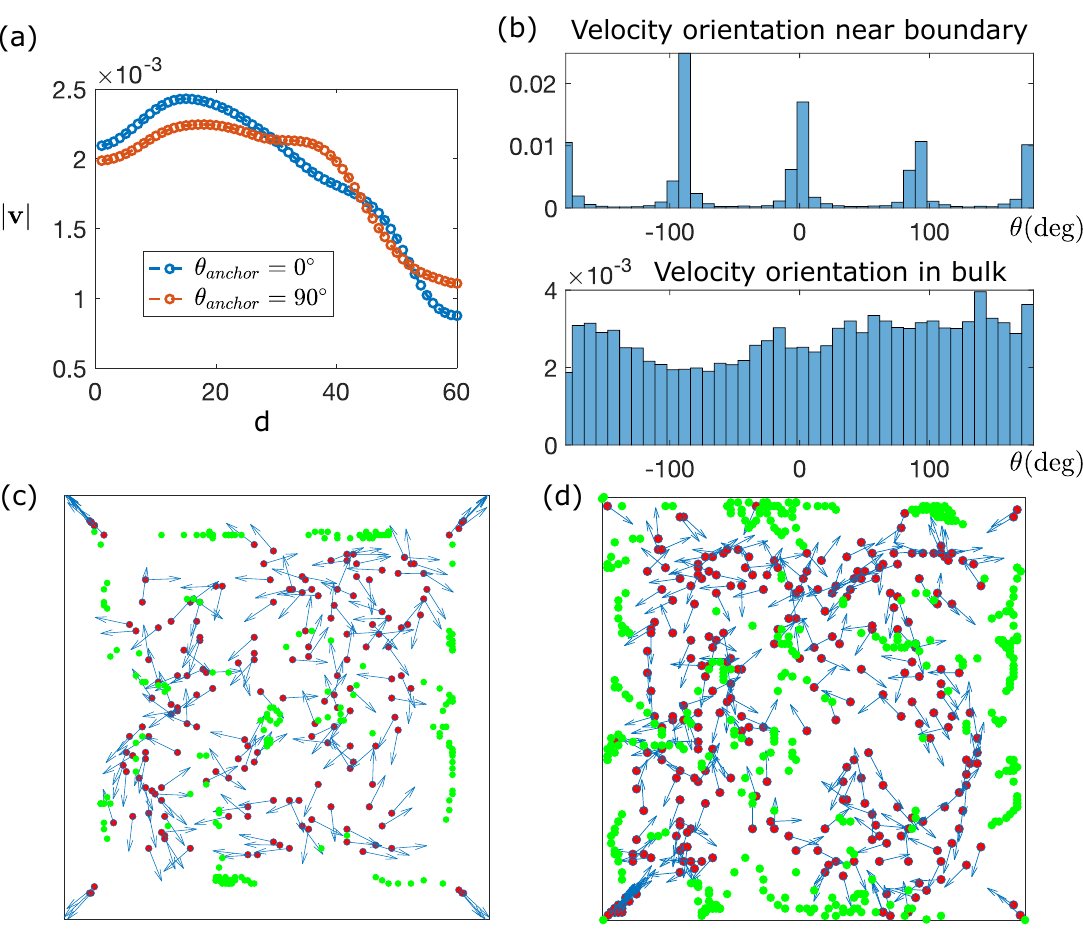}
    \caption{\textbf{Velocity field and defects in the bulk and boundaries:} \textit{(a)} The average velocity of the fluid is higher in the bulk than the boundaries. \textit{(b)} Orientation of the velocity field in a square-annulus near the boundary and inside the bulk. Flows are more uniform along the boundary, where they line up along the walls. In the bulk, flows are turbulent and less uniform. \textit{(c)} \textbf{Extensile activity, planar anchoring:} Defect locations for planar anchoring across all time snapshots. Red dots denote $+1/2$ defects, and blue arrows denote the direction of defect motion. Green dots denote $-1/2$ defects. \textit{(d)} \textbf{Extensile activity, homeotropic anchoring:} Defect locations for homeotropic anchoring across all time snapshots. Defects tend to point towards (away from) the corners in the planar (homeotropic) case.}     \label{fig:DefVel}
\end{figure}

\begin{figure}[H]
    \centering
    \includegraphics[width = 0.75\textwidth]{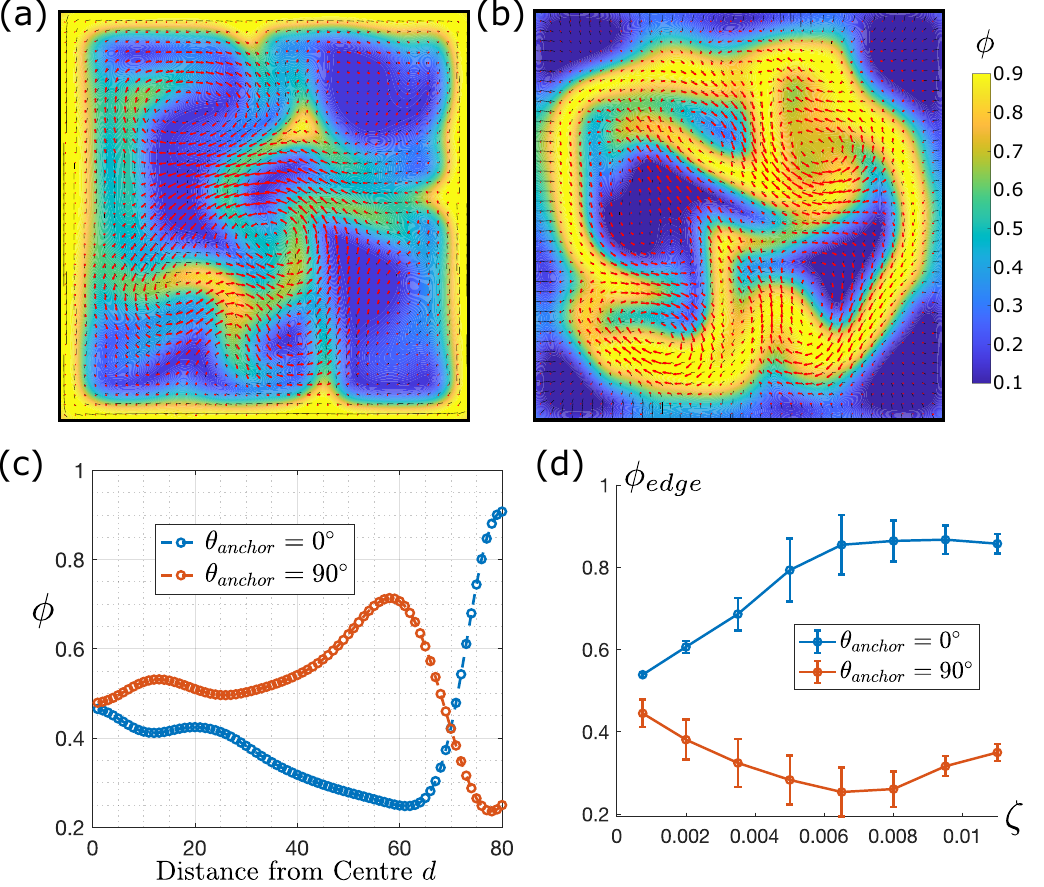}
    \caption{\textbf{Active nematic-passive mixture with imposed strong anchoring in a no-slip box:} \textit{(a)} \textbf{Extensile activity, planar anchoring:} For extensile activity ($\zeta=0.008$), with planar anchoring, the concentration of active nematic at the boundary wall is enhanced. The colourbar shows the concentration of active nematic fluid $\phi$. Black lines denote the orientation of the nematic field, and red arrows denote the velocity field. \textit{(b)} \textbf{Extensile activity, homeotropic anchoring:} If the anchoring is changed to homeotropic, the concentration of active nematic at the boundary is depleted. \textit{(c)} Variation of active concentration as a function of distance $d$ from the centre of the box. Averages are taken over square-annuli of side 2$d$. \textit{(d)} Concentration of active nematic near the wall, averaged over time, for varying activity. Error bars show the standard deviation.}     \label{fig:NoSlip}
\end{figure}

\begin{figure}[htp]
    \centering
    \includegraphics[width = 0.8\textwidth]{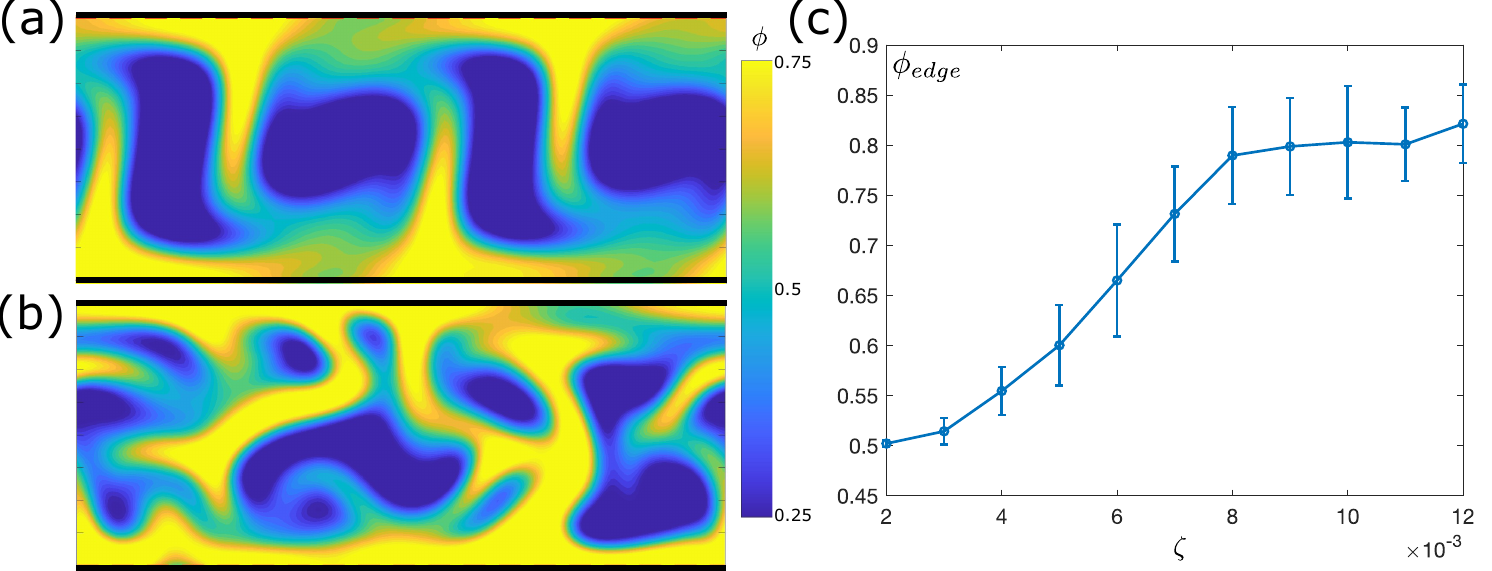}
    \caption{\textbf{Spontaneous sorting in a no-slip channel.} \textit{(a)} \textbf{\textbf{Extensile activity, active anchoring:}} Concentration of the active component when the bulk forms a vortex lattice ($\zeta = 0.006$). Colourbar shows the concentration of the active component, $\phi$. \textit{(b)} \textbf{\textbf{Extensile activity, active anchoring:}} Concentration of the active component in a turbulent configuration ($\zeta = 0.012$). \textit{(c)} Variation of active concentration at the edge of the channel with increasing activity.}     \label{fig:Channel}
\end{figure}

\begin{figure}[htp]
    \centering
    \includegraphics[width = \textwidth]{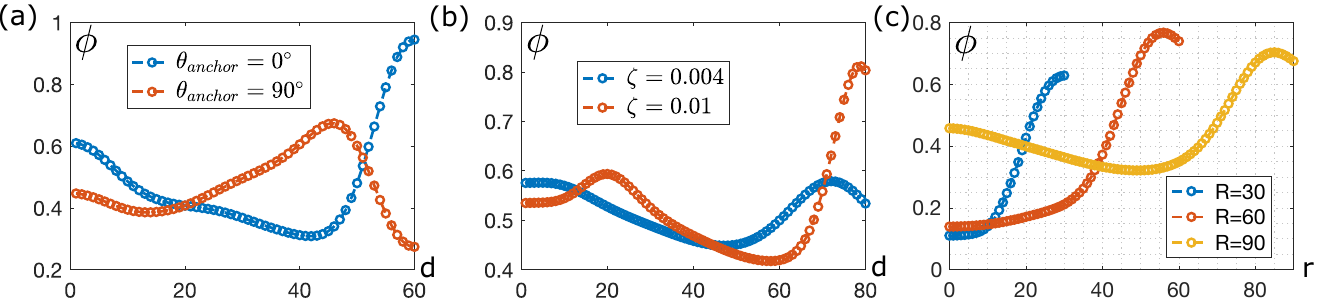}
    \caption{\textbf{Spatial variation of the active concentration field $\phi$:} \textit{(a)} For extensile activity - planar (homeotropic) anchoring results in enhanced (depleted) concentration of active nematic at the boundary. Blue and red data points correspond to Figs.~1a--b (in the main text) respectively. $d$ denotes the distance from the centre of the box. $\phi$ is averaged using square-annuli. \textit{(b)} Active sorting to the boundary due to active anchoring, shown for different activities. Blue and red data points correspond to Figs.~2a--b respectively. \textit{(c)} Spatial variation of the active concentration in circular confinement as a function of radial distance $r$. Blue, red and yellow data points correspond to Figs.~3a--c respectively.}     \label{fig:Phi_CrossSection}
\end{figure}

\end{document}